\documentclass[11pt]{article}

\usepackage[T1]{fontenc} 
\usepackage{microtype} 
\usepackage{amssymb,amsmath,amsthm,amscd}

\newtheorem{corollary}{Corollary}

\newtheorem{proposition}{Proposition}

\usepackage{geometry}
\usepackage{paralist}
\usepackage{longtable}
\usepackage[titletoc,toc,title]{appendix}
\usepackage{booktabs} 
\usepackage{float} 
\usepackage{hyperref} 
\usepackage[utf8]{inputenc}
\usepackage{dsfont}
\usepackage{lettrine} 
\usepackage{paralist} 
\usepackage{abstract} 
\usepackage{color}
\usepackage{graphicx}
\usepackage{mathrsfs}
\usepackage{epstopdf}
\numberwithin{equation}{section}
\begin{document}
\title{Rough volatility: evidence from range volatility estimators}

\author{Saad Mouti\\ University of California, Santa Barbara\\ saadmouti@ucsb.edu}
\date{%
{\parbox{\linewidth}{\centering%
  November 1, 2023\thanks{This paper is part of the thesis \cite{mythesis} co-supervised by Mathieu Rosenbaum, Nicole El Karoui, and Aymeric Kalife, and benefited from funding provided by AXA. I would like to thank Thibault Jaisson, Omar El Euch, Pamela Saliba, and Othmane Mounjid for their valuable discussions. My appreciation also extends to participants at the Imperial College Rough Volatility symposium, UC Berkeley's CDAR seminar, and the 2023 Southern California Probability Symposium at UC San Diego for their constructive feedback.} \endgraf}}
}
\maketitle

\noindent \textit{\textbf{Abstract---}}
{\noindent In \cite{GJR141}, first posted in 2014, volatility is characterized by fractional behavior with a Hurst exponent $H < 0.5$, challenging traditional views of volatility dynamics. Gatheral et al. demonstrated this using realized volatility measurements. Our study extends this analysis by employing range-based proxies to confirm their findings across a broader dataset and non-standard assets. Notably, we address the concern that rough volatility might be an artifact of microstructure noise in high-frequency return data. Our results reveal that log-volatility, estimated via range-based methods, behaves akin to fractional Brownian motion with an even lower $H$, below $0.1$. We also affirm the efficacy of the rough fractional stochastic volatility model (RFSV), finding that its predictive capability surpasses that of AR, HAR, and GARCH models in most scenarios. This work substantiates the intrinsic nature of rough volatility, independent of the microstructure noise often present in high-frequency financial data.}
\medskip
\par\noindent \textbf{\textrm{Keywords}}: Range-based volatility; Garman-Klass; Parkinson; volatility scaling; fractional Brownian motion; fractional Ornstein-Uhlenbeck; volatility forecasting.

\section{Introduction}
\noindent Volatility plays a crucial role in many areas of finance and economics, such as risk management and portfolio selection. It is known to be both time-varying and predictable, and stochastic volatility models are one way to deal with these features. As a consequence, its modeling and forecasting spur the interest of many authors, academics, and practitioners alike.\\
\\
\noindent In the financial markets, the common practice is to represent asset prices by a continuous semi-martingale. A given log-price $\log S_t$ is defined by
\begin{align*}
d\log S_t = \mu_tdt + \sigma_t dW_t
\end{align*}
where $\mu_t$ is a drift term, $W_t$ a Brownian motion and $\sigma_t$, the key ingredient is the process volatility. The Black-Scholes framework assumes the volatility to be either constant or deterministic. Such specifications proved to be inadequate in the late eighties. The main reason is the inconsistency of the Black-Scholes model with the observed European options. This gave rise to alternatives such as local volatility models, such as Dupire's, see \cite{dupire1994pricing}, and Derman and Kani's, see \cite{derman1994riding}. These models consider $\sigma_t$ as a deterministic function of time and asset price. Even though they enable us to perfectly fit a given implied surface, its dynamic is quite unrealistic. Another alternative is to model the volatility $\sigma_t$ by a continuous Brownian semi-martingale, typically correlated with $W$. These so-called stochastic volatility models have been the center of interest of many authors. We cite amongst such stochastic volatility models, the Hull and White model \cite{hull1987pricing}, the Heston model \cite{heston1993closed}, and the SABR model \cite{hagan2002managing}. However, generated option prices are still not consistent with observed European option prices. The reader can refer to \cite{gatheral2006} for a review of different approaches. More recently, the market practice uses so-called local-stochastic volatility models, which fit the market exactly and generate reasonable dynamics.\\
\\
Since volatility is a latent variable, the first issue one faces when trying to exhibit its statistical properties is its estimation. One can only estimate it using the underlying asset prices or quoted options. For example, when we have daily stock returns, the squared variance is a well-known volatility proxy known as realized volatility. It measures changes in the asset return over a specified period of time. If high-frequency data is available (the whole price process during the day), this proxy is more precise and is used as the daily volatility.\\
\\
Access to high-frequency data is sometimes costly and/or unavailable for certain assets. Therefore, other proxies are used to estimate daily volatility. If we only have closing prices and need to estimate volatility on a daily basis, we can use the squared daily returns. A compromise can be found as, in addition to closing prices, open, high, and low daily prices are available for most financial data sets. In \cite{Park801}, Parkinson was the first to introduce an advanced volatility estimator using these so-called range prices instead of just closing prices, and this enables us to overcome the issues of the first two approaches and present a way to better estimate daily volatility.\\
\\
Statistical properties of volatility estimators raise interesting questions, particularly in relation to the smoothness of the volatility process. Researchers aim to uncover the underlying mechanisms that generate the data, using the empirical scaling evidence as a stylized fact that any theoretical model should also reproduce. For example, it is a common belief that volatility exhibits what is commonly known as long-range dependence. The implication of this is that volatility shocks today will influence its expectation in the same direction, see \cite{BOST131,EP011} among others.\\
\\
Stochastic or local volatility models mentioned earlier assume that the smoothness of the sample path of the volatility is that of a Brownian motion ($1/2 - \varepsilon$ H\"older continuous for any $\varepsilon > 0$). In \cite{CR981}, Comte and Renault choose to address the question of long-range dependence in terms of the regularity of the driving process. Their idea was to exploit the fractional Brownian motion. Recall that a fractional Brownian motion $W^H$ with Hurst parameter $H\in (0,1)$ is an a.s. continuous, centered, self-similar Gaussian process with stationary increments and a covariance satisfying:
\begin{equation*}
Cov(W_t^H,W_s^H) = \frac{1}{2}(|t|^{2H}+|s|^{2H}-|t-s|^{2H}), t, s \in \mathbb{R}
\label{equ1}
\end{equation*}
They proposed the fractional Ornstein-Uhlenbeck volatility model with a Hurst parameter greater than $1/2$ named fractional volatility model (FSV). Such a model is $H-\varepsilon$ H\"older continuous with $H > 1/2$. More interest grew from this model, and others developed deeper analysis and calibration, see \cite{CKM031,CCER121} among others. Later, Gatheral and co-authors, see \cite{GJR141}, challenged the previous results and established that the log-volatility process is very close to that of a fractional Brownian motion with Hurst parameter around $0.1$ ($<1/2$). They also developed the rough fractional stochastic volatility model (RFSV), which operates with different parameter properties than the FSV, and justified that their model better respects the volatility smile and the data properties.\\
\\
Recent discussions in the field have raised questions about the rough volatility paradigm, particularly from a statistical perspective. While some skepticism exists, several studies, such as those by \cite{bennedsen2022decoupling}, and \cite{fukasawa2019volatility}, have confirmed the roughness of volatility paths. Although alternative models can match the effectiveness of rough volatility in specific tasks, they often lack its parsimony. Various volatility proxies and a range of tests strongly suggest that the rough nature of volatility is not a mere artifact and is challenging to refute. A key statistical question then emerges: how to infer the Hurst parameter $H$. The ability to accurately estimate $H$ and differentiate between its values is crucial to overcome the latent nature of volatility and the noise in its estimation. Recent research, including work by \cite{chong2022statistical_clt}, has achieved an optimal speed of convergence for estimating $H$ using wavelets, and \cite{chong2022statistical_minimax} have developed a consistent estimator of $H$ based on price observations. Additionally, \cite{BOLKO2022} introduced a Generalized Method of Moments (GMM) approach to estimate the roughness of stochastic volatility.\\
\\
In this paper, we extend the methodologies used in \cite{GJR141, livieri2018rough} to analyze range-based estimators, thereby exploring the rough volatility hypothesis to a broader spectrum of data. This expansion enables us to revisit previous findings focusing on non-standard assets, thanks to the increased availability of data. Our results reveal that the Hurst parameter for range-based proxies within our dataset is notably lower than $0.1$, and, in some cases, even approaches $0$. This strongly supports the rough volatility hypothesis. Moreover, our analysis provides sufficient grounds to dismiss the fractional stochastic volatility (FSV) model, which posits a Hurst parameter greater than $0.5$.\\
\\
Our paper is organized as follows. In Sections \ref{Sect1.2}, we give an overview of range-based volatility estimation before choosing one volatility proxy to work with. We conduct our statistical study in Section \ref{Sect1.3}, finding that log-volatility from range-based proxies behaves like a fractional Brownian motion with Hurst exponent lower than $0.1$. We validate the rough fractional stochastic volatility model (RFSV) introduced in \cite{GJR141} on our data in Section \ref{Sect1.4}, compare its prediction power with other common models in Section \ref{Sect1.5}, and finally conclude in Section \ref{Sect1.6}.
\section{Overview on range-based volatility estimation}\label{Sect1.2}
\begin{figure}[ht]
\centering
\includegraphics[width=16cm,height=8cm]{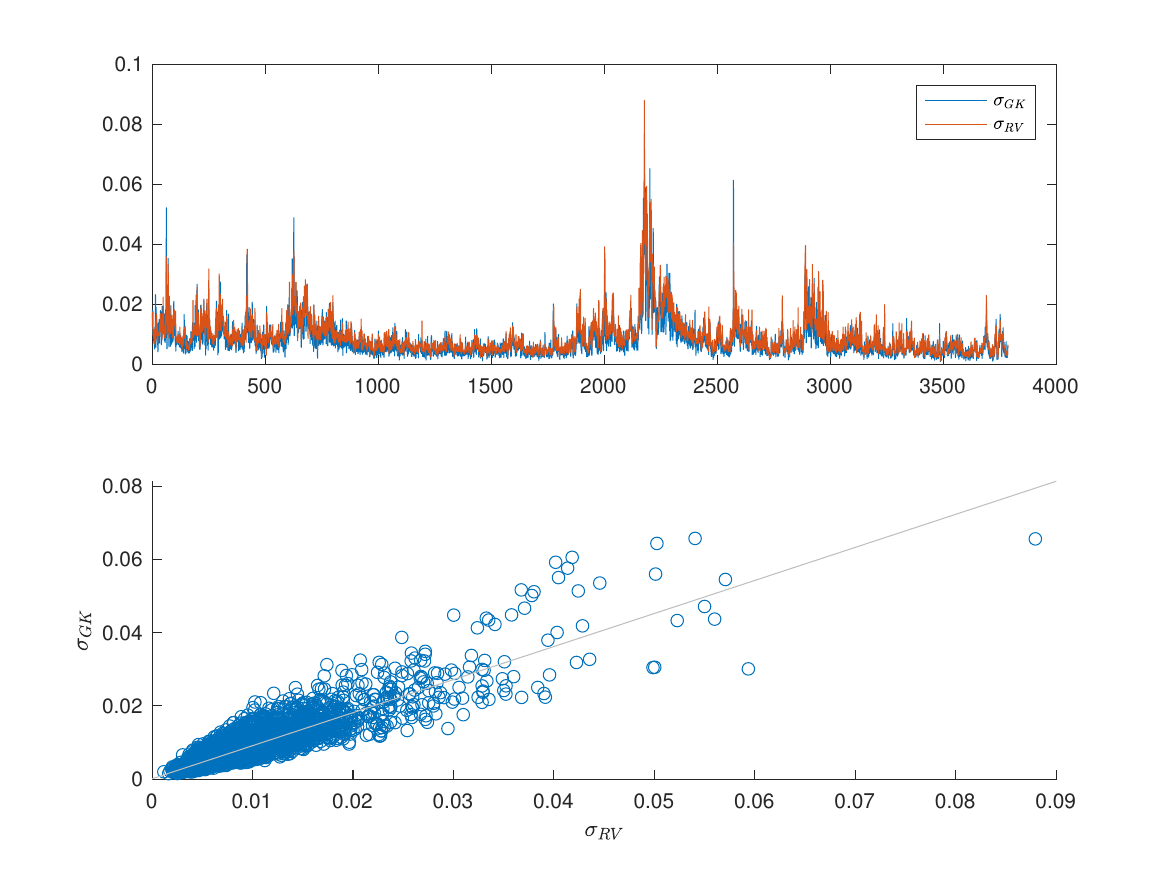}\\
\caption{\small{Comparison between the GK proxy and the RV proxy as a benchmark. Top is the evolution of the volatility time series for both estimators. The bottom graph is a scatter plot of the GK estimator to the RV estimator.}}
\label{figCompare0}
\end{figure}

\begin{figure}[ht]
\centering
\includegraphics[width=16cm,height=10cm]{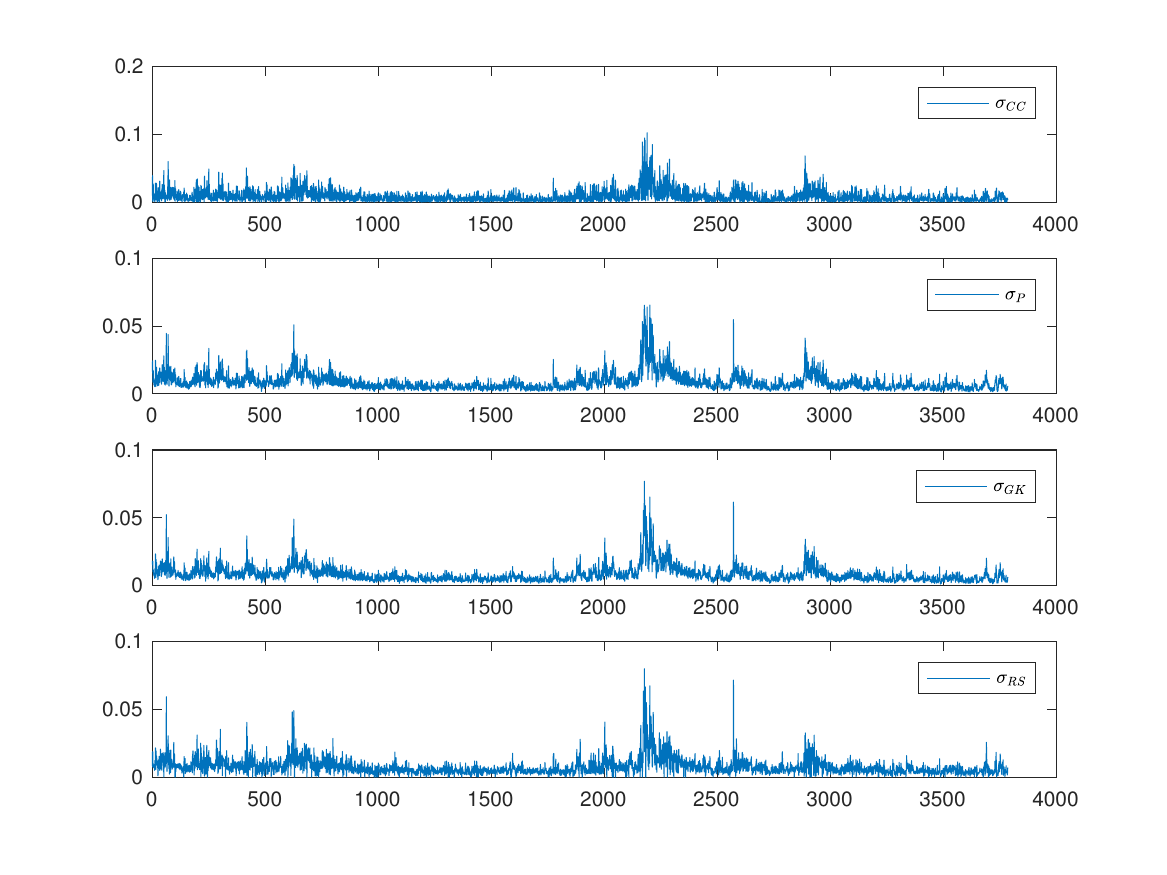}
\caption{\small{Comparison between different range-based estimators.}}
\label{figCompare1}
\end{figure}

\noindent In this section, we review a few range-based volatility estimators and compare them to realized volatility.\\ 
\\
We assume that the asset price over a one-day period of time, $S_u$, follows a geometric Brownian motion:
\begin{equation*}
dS_u  = \mu S_udu + \sigma S_u dW_u,
\end{equation*}
where $u \in (t-1,t]$ is the time index between two consecutive days $t-1$ and $t$, $\mu$ is the drift, $\sigma$ the volatility considered constant along one day, and $W_u$ a standard Brownian motion. By Ito's lemma the log price $\log(S_u)$ follows a Brownian motion with drift $\mu^* = \mu - \frac{\sigma^2}{2}$ and volatility $\sigma$. During a day, it is common practice to assume that the drift is equal to zero (i.e., $\mu^*\approx 0$).\\ 
\\
Volatility is a latent variable, and one needs to estimate it at each period of time with the available asset prices. We denote by $H$, $L$, $O$, $C$ the high, low, open, and close prices, respectively. The log-returns $r_t$ are defined by $r_t = \log(C_t) - \log(O_t)$ (or $\log(C_t) - \log(C_{t-1})$ when taking only close-to-close prices). Volatility changes and our first interest is to be able to estimate it in a precise way and on a daily basis. Taking the squared return $r_t^2$ is one possible solution. It is an unbiased estimator of $\sigma^2$ under the normal log-returns assumption with zero mean. However, this estimator is quite noisy.\\
\\
Range prices bring more consistency and information about the entire process than the close-to-close prices. In this context, Beckers, see \cite{Bec831}, shows that volatility estimators can be improved by incorporating high and low prices and closing prices. Of course, range-based volatility estimators are not as efficient as realized volatility under ideal conditions, i.e., estimated from high-frequency data, but remain a good alternative when this data is not available. It was shown in \cite{ABD021} and \cite{SZ061} that these range-based estimators are robust to microstructure noise and prove to be efficient and simple to compute.\\
\\
In \cite{Park801}, Parkinson was the first to develop a classical range-based estimator using high and low price information expressed by the following formula:
\begin{equation*}
\sigma_{\text{Parkinson}}^{2} = \frac{1}{4\log 2}\Big(\log\frac{H_t }{L_t}\Big)^2 .
\end{equation*}
The correctness of this formula relies on a constant volatility assumption during each one-day time period. It exploits the extreme value method, and the coefficient $\frac{1}{4\log 2}$ is nothing but the variance of the range variable (i.e., the difference between the minimum and maximum). The reader can refer to \cite{Park801} for details and to \cite{Feller51} for the asymptotic distribution of the range.\\
\\
The Parkinson estimator is asymptotically unbiased under the assumption that a geometric Brownian motion without drift can describe the path of the asset price changes.\\
\\
Later, Garman and Klass, see \cite{GK801}, established a more efficient estimator that takes the following form:
\begin{align*}
\sigma_{GK}^2 = 0.511\left(\log\frac{H_t}{L_t}\right)^2 - 0.019&\left(\log\frac{C_t}{O_t}\left(\log\frac{H_t}{O_t}-\log\frac{L_t}{O_t}\right) \right. & \nonumber \\
& \left. - 2\log\frac{H_t}{O_t}\log\frac{L_t}{O_t}\right) - 0.383\left(\log\frac{C_t}{O_t}\right)^2. &
\end{align*}
\noindent This estimator combines the squared return and Parkinson volatility estimators into a new estimator with a smaller variance. Garman and Klass actually proved that this estimator is optimal in a mean-variance sense among a certain class of estimator, see \cite{GK801}.\\
\\
A more practical estimator is recommended with nearly the same efficiency but eliminates the small cross-product terms expressed as:
\begin{equation}
\sigma_{GK}^2 = \frac{1}{2}\left(\log\frac{H_t}{L_t}\right)^2 - (2\log2-1)\left(\log\frac{C_t}{O_t}\right)^2 .
\end{equation}
We rely on this estimator when using the Garman-Klass volatility proxy.\\
\\
Because in \cite{GK801} and \cite{Park801} log prices are assumed to follow geometric Brownian motion with no drift, many authors tried to correct this mismatch for securities with non-zero mean. In \cite{RS911}, more sophisticated drift-independent measures of volatility are introduced. The Rogers-Satchell estimator, for example, takes the form:
\begin{equation*}
\hat{\sigma}^2_{RS} = \log\frac{H_t}{O_t}\left(\log\frac{H_t}{O_t}- \log\frac{C_t}{O_t}\right) + \log\frac{L_t}{O_t}\left(\log\frac{L_t}{O_t} - \log\frac{C_t}{O_t}\right) .
\end{equation*}
Kunimoto \cite{Kun921} and Yang-Zhang \cite{YZ001} also deserve to be mentioned. Yang-Zhang estimator, however, can only be used over multiple days and therefore won't be interesting for our analysis.\\
\\
The previously mentioned estimators are unbiased estimators of $\sigma^2$. When applying the square root and estimating the volatility, all $\sigma$ estimators are biased. This was expected since $\mathbb{E}[\sigma^2]$ and $\mathbb{E}[\sigma]^2$ are generally different.\\
\\
In terms of efficiency, all previous estimators exhibit very substantial improvements compared to the close-to-close estimator. The efficiency measure of a volatility estimator $\hat{\sigma}_i^2$ is defined as the ratio of the variance of this estimator and the variance of the close-to-close estimator $\hat{\sigma}_{CC}^2$:
\begin{align*}
\text{Eff}(\hat{\sigma}_t^2) = \frac{\mathbb{V}ar(\hat{\sigma}^2_{CC})}{\mathbb{V}ar(\hat{\sigma}^2_i)}.
\end{align*}
By definition, the squared return estimator has an efficiency of $1$. Parkinson reported that his estimator is 2.5 to 5 times more efficient than a simple close-to-close variance estimator. Garman-Klass reports 7.4 while Rogers-Satchell efficiency is 6.0 and Kunimoto is 10.\\
\\
In light of \cite{BW051}, we compare the performance and distributional properties of different range-based volatility estimators on the S\&P 500 over a 3786 period from January 2000 to April 2015. Our benchmark is the realized volatility from the Oxford-Man Institute of Quantitative Finance Realized Library. To perform the comparison, we use the following measures: \\
The mean squared error is defined by:
\begin{align*}
MSE(\sigma_{\text{estimated}}) = \mathbb{E}\left[(\sigma_{\text{estimated}}-\sigma_{\text{benchmark}})^2\right].
\end{align*}
The mean absolute bias is given by:
\begin{align*}
MAD(\sigma_{\text{estimated}}) = \mathbb{E}\left[\mid\sigma_{\text{estimated}}-\sigma_{\text{benchmark}}\mid\right].
\end{align*}
The proportional bias is expressed as:
\begin{align*}
\text{Prop.Bias}(\sigma_{\text{estimated}}) = \mathbb{E}\left[(\frac{\sigma_{\text{estimated}}}{\sigma_{\text{benchmark}}}-1)\right].
\end{align*}
Recall that the benchmark volatility $\sigma_{\text{benchmark}}$ here corresponds to realized volatility.\\
\\
The results are given in Figures \ref{figCompare0} and \ref{figCompare1} and Table \ref{TabComp}.
\begin{center}
\begin{longtable}{| p{.15\textwidth} | p{.15\textwidth} | p{.15\textwidth} | p{.15\textwidth}| p{.15\textwidth} |} 
\hline
 & $MSE$ & $MAD$ & $\text{Prop.Bias}$ & Std.Dev \\
\hline
$\hat\sigma_{CC}$   &  $0.533\times 10^{-4}$  &  0.0051 &  -0.1275 & 0.0092 \\
\hline
$\hat\sigma_{P}$  &  $0.092\times 10^{-4}$  &  0.0021 &  -0.0849 & 0.0063 \\
\hline
$\hat\sigma_{GK}$  &  $0.094\times 10^{-4}$  &  0.0021  & -0.1313 & 0.0058  \\
\hline
$\hat\sigma_{RS}$  &  $0.206\times 10^{-4}$ &   0.0028  & -0.1762 & 0.0062  \\
\hline
\caption{\small{Comparison measures for different volatility estimators}}
\label{TabComp}
\end{longtable}
\end{center}
We can see in Table \ref{TabComp}, Figure \ref{figCompare0} and Figure \ref{figCompare1} the following:
\begin{itemize}
	\item Range-based estimators are lower than the benchmark.
	\item Range-based estimators reduce the variance compared to squared returns.
	\item The Garman-Klass estimator is the closest to intraday realized variance and has the smallest variance.
	\item Range-based estimators seem to exhibit the same regularity as that of realized volatility.
\end{itemize}
\noindent We would like to confirm the last point. In our study, we focus on the volatility based on the Garman-Klass proxy. Our goal is to confirm that the scaling properties discussed in \cite{GJR141}, are also satisfied by the volatility when using these proxies.
\section{Range-based volatility as spot volatility proxy: empirical results}\label{Sect1.3}
\subsection{The scaling of the Garman-Klass proxy}
\begin{figure}[ht]
\centering
\includegraphics[scale=0.45]{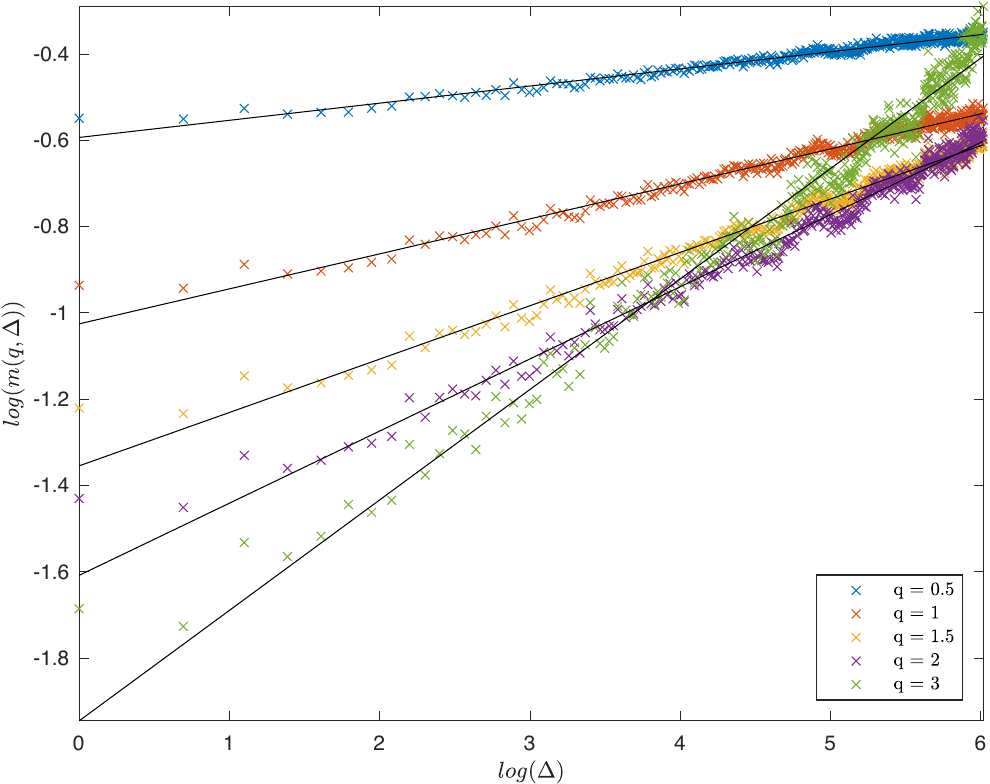}
\includegraphics[scale=0.45]{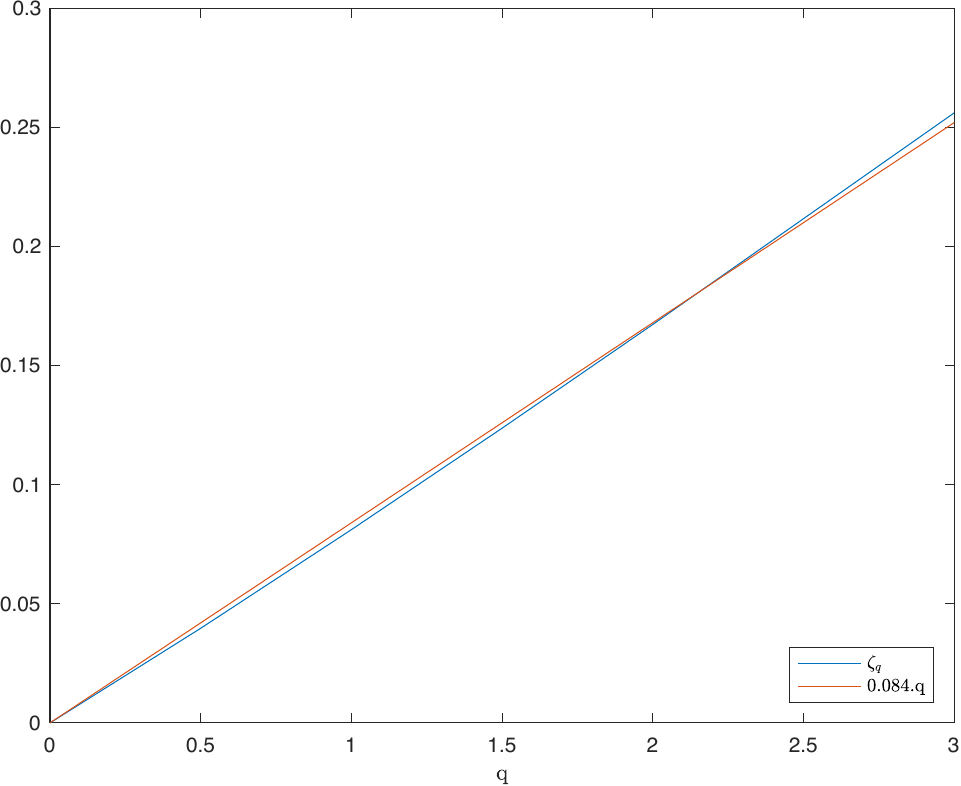} 
\caption{$\log m(q,\Delta)$ as a function of $\log \Delta$ (left), $\zeta_q$ (blue) and $0.084 \times q$ (green) (right), S\&P 100 (Garman Klass volatility).}
\label{fig1SP100}
\end{figure}

\begin{figure}[ht]
\centering
\includegraphics[scale=0.45]{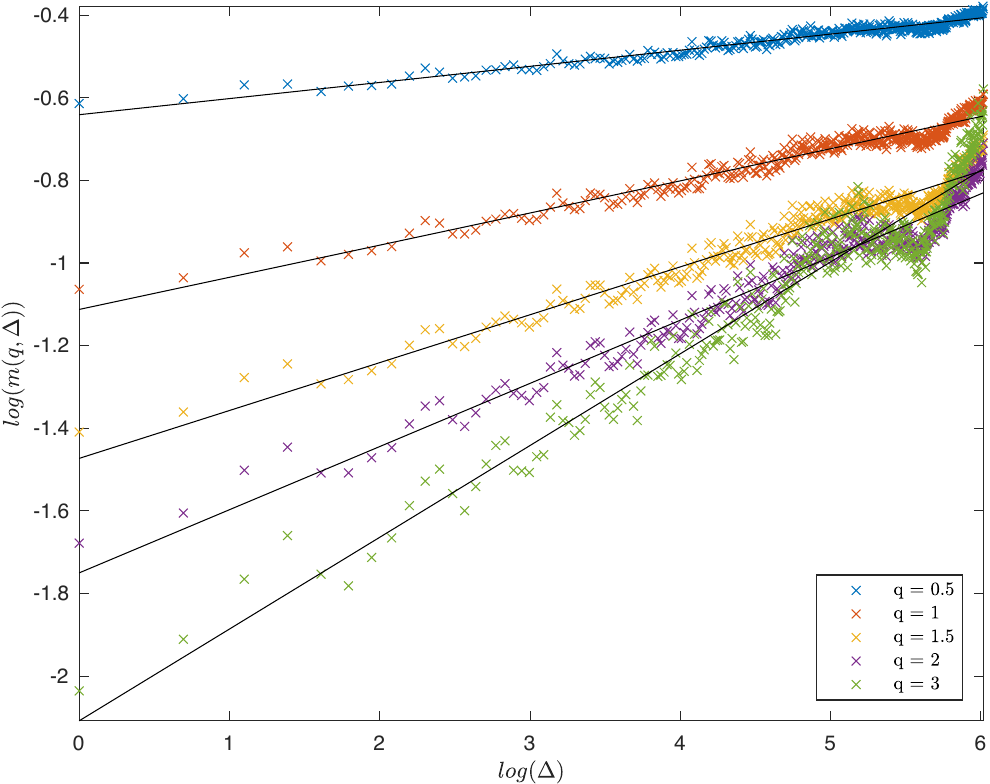}
\includegraphics[scale=0.45]{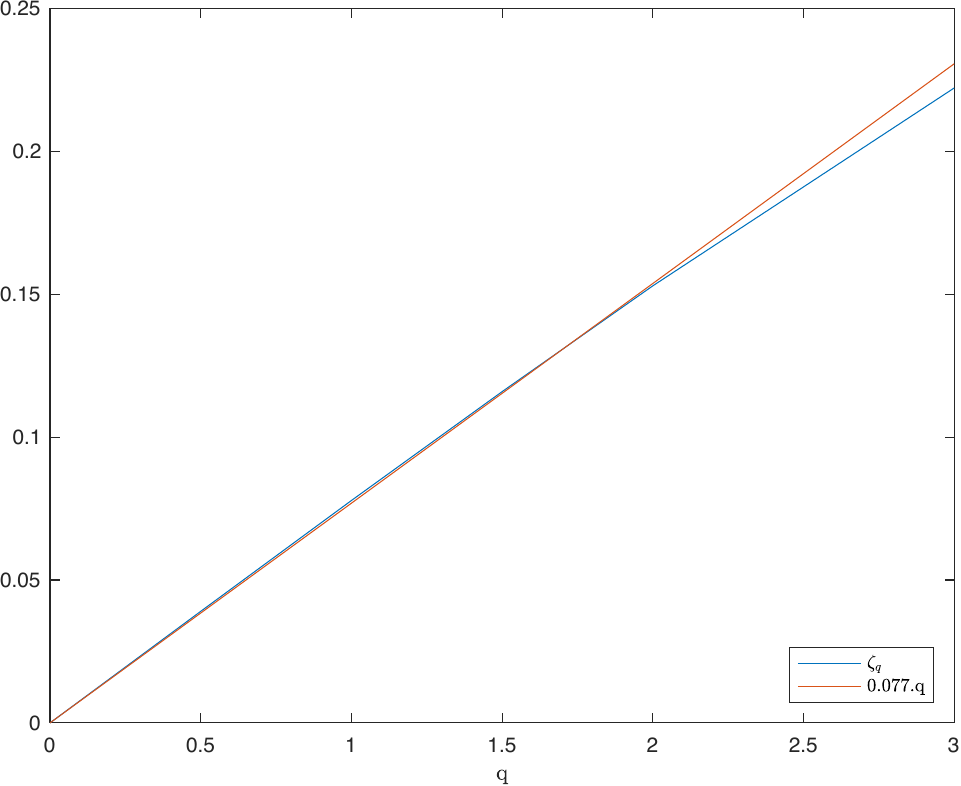} 
\caption{$\log m(q,\Delta)$ as a function of $\log \Delta$ (left), $\zeta_q$ (blue) and $0.072 \times q$ (green) (right), IBEX35 (Garman Klass volatility).}
\label{fig1IBEX35}
\end{figure}

\noindent We carry out our analysis on the volatility proxy for a variety of assets. In \cite{GJR141}, Gatheral and co-authors use common indexes for which high frequency based realized volatility is available on the Oxford-Man Institute of Quantitative Finance Realized Library (S\&P 500, Bund ...). In this paper, we choose to apply our analysis on more "exotic" assets (S\&P 400, IBEX 35, IBOV, S\&P 100, INDU, SHSZ300, MEXBOL, FTSE 100, XIN9I, HSI), and some stocks (TOTAL, ASX200, GOOGLE and MICROSOFT). Most of these assets are not available on the Library, and more importantly, since many financial institutions still do not have access to high frequency based proxies, the choice of range-based is very convenient. Range data availability allows for their computation for any class of assets and any assets tickers. \\
\\
We choose to present the analysis for the S\&P 100\footnote{The S\&P 100 is a subset of the S\&P 500 and includes 102 leading U.S. stocks with exchange-listed options. Constituents of the S\&P 100 represent about 63\% of the market capitalization of the S\&P 500 and almost 51\% of the market capitalization of the U.S. equity markets as of January 2017.} and IBEX 35\footnote{IBEX 35 is the benchmark stock market index of the Bolsa de Madrid, Spain's principal stock exchange. It is a market capitalization weighted index comprising the 35 most liquid Spanish stocks traded in the Madrid Stock Exchange General Index}, and give numerical results for the remaining ones. We focus mainly on the Garman-Klass proxy and present results on the Parkinson estimator for verification.\\
\\
The set of data corresponds to 2521 trading days from April 19, 2005 to April 22, 2015. Volatility proxies are based on range data extracted from the Bloomberg database. Let $\sigma_{t_0}, \sigma_{t_1},...,\sigma_{t_N}$, be the time series of the GK proxy computed for this period, where $t_{i+1} - t_i$ corresponds to one business day.\\
\\
Our scaling measure $m(q,\Delta)$ is the $q$-th absolute moment of the increments of the log-volatility, defined by:
\begin{align*}
m(q,\Delta) := \frac{1}{N}\sum_{k = 1}^{\lfloor N/\Delta \rfloor}{\mid \log(\sigma_{k\Delta}) - \log(\sigma_{(k-1)\Delta}) \mid^q},
\end{align*}
for different $q > 0$ and lags $\Delta$ going from 1 to about 400 days. Our goal is to revisit the finding in \cite{GJR141} that the spot log-volatility has the same scaling properties as a fractional Brownian motion with Hurst exponent $H < 1/2$, and therefore, one can model it with such process. It is worth to mention the following remarks: 
\paragraph{Remarks:}
	\begin{itemize}
		\item The quantity $m(q,\Delta)$ is the discrete equivalent of $\mathbb{E}[\mid \log(\sigma_\Delta) - \log(\sigma_0) \mid^q]$. Recall that the fractional Brownian motion $W^H$ verifies:
		\begin{align*}
		\mathbb{E}[\mid W^H_{t+\Delta} - W^H_t \mid^q] = \tilde{K}_q \Delta^{qH},
		\end{align*}
		\item We are expecting the volatility to behave closely to the fractional Brownian motion. As a result, we would observe the following relationship:
		\begin{align}\label{scalingLogvol}
		m(q,\Delta) \sim K_q \Delta^{qH}.
		\end{align}
		\item Given that data is finite and not time-equidistant (unavailable data on weekends, for example), the measure $m(q,\Delta)$ is the result of averaging over all possible increments by taking a rolling window and selecting only increments that correspond to the chosen lag $\Delta$ (log-vol increments between two successive volatility measures between Friday through Monday are considered as three days lag).
	\end{itemize}
\noindent To verify the validity of (\ref{scalingLogvol}), we plot $\log(m(q,\Delta))$ against $\log(\Delta)$ for different values of $q$. Depending on the results of this first regression, we can write, for a given $q$,
\begin{align}\label{eq:genScaling}
m(q,\Delta) \sim K_q \Delta^{\zeta_q},
\end{align}
where $\zeta_q$ defines a general scaling function.\\
\\
Results are displayed in Figures \ref{fig1SP100} and \ref{fig1IBEX35}. We can notice the following:
\begin{itemize}
	\item The values of $m(q,\Delta)$ for different $q$ against $\log(\Delta)$ lie within a straight line (left figures). This confirms that both S\&P 100 and IBEX 35 exhibit a scaling property given by Equation (\ref{eq:genScaling}).
	\item The R-squared values given in Table \ref{tab:RSquared} confirm that the data is close to the fitted regression line for all values of $q$. 
 	\item Plotting $\zeta_q$ as a function of $q$ (right), confirms our expectation; the scaling is linear in $q$ and verifies Equation (\ref{scalingLogvol}), with $H = 0.081$ for S\&P 100 and $H = 0.072$ for IBEX 35.
\end{itemize}
\begin{table}[H]
\centering
\begin{tabular}{llllll}
\hline 
 $q$ & $0.5$ & $1$ & $1.5$ & $2$ & $3$  \\ 
\hline 
S\&P 100 & $0.9640$ &   $0.9760$  &  $0.9800$  &  $0.9797$  &  $0.9717$ \\
IBEX 35 & $0.9225$  &  $0.9290$  &  $0.9287$  &  $0.9252$  &  $0.9111$ \\
\hline 
\end{tabular}
\caption{The R-squared measure of the regression $\log(m(q,\Delta)) \sim \log(\Delta)$}
\label{tab:RSquared}
\end{table}
\noindent Of course the $\zeta_q$ and $Hq$ are not perfectly matched. One possible reason for this mismatch is using discrete samples. In fact, simulating fBm using the same number of points results in a slightly concave figure and ends up in recovering a Hurst parameter that slightly overestimates the real one.\\
\\
\noindent In order to make sure our estimations of $H$ do not depend on the time interval, we split the data into two periods with the same lag and re-estimate $H$ for each period separately. The aim of this experience is to confirm that the scaling is time-independent for all assets. Regressing $\log(m(q,\Delta))$ on $\log(\Delta)$ for each ticker and for $q = 0.5, 1, 1.5, 2, 3$ and $\Delta = 1,...,410$, we find that $\zeta_q$ is linear on $q$ for the GK proxy. As we can see in Table \ref{Tab1}, the Hurst parameter lies between $0.01$ and $0.082$, confirming that the volatility process is rough. One might, however, notice that splitting the data resulted in a first half with a slightly greater $H$ than the second half. We think that it is due to the presence of the 2008 crisis in the first half.
\begin{table}[H]
	\centering
	\begin{tabular}{cccc}
Ticker & $H$ for the whole period & $H$ (first half) & $H$ (second half) \\
\hline
SP100 &0.0841 &0.0897 &0.0714\\
IBEX35 &0.072 &0.0753 &0.071\\
HSI &0.0516 &0.0605 &0.0394\\
MEXBOL &0.0627 &0.0737 &0.0463\\
FTSE100 &0.0751 &0.0708 &0.0728\\
ASX200 &0.0489 &0.0476 &0.0415\\
TOTAL &0.0774 &0.0835 &0.0687\\
XIN9I &0.0674 &0.0649 &0.069\\
SHSZ300 &0.0689 &0.0718 &0.0636\\
BCOM &0.014 &0.0099 &0.0238\\
INDU &0.0804 &0.0838 &0.067\\
USDEUR &0.0353 &0.0393 &0.0321\\
IBOV &0.0685 &0.0724 &0.0609\\
MICROSOFT &0.06 &0.0717 &0.0401\\
GOOGLE &0.0656 &0.0724 &0.0542\\
SP400 &0.0715 &0.0753 &0.0592\\
\hline
\end{tabular}

	\label{Tab1}
	\caption{Estimates of $H$ on the whole period and over two different time intervals for different indexes and stocks (Garman Klass volatility)}
\end{table}
\subsection{The scaling of the Parkinson volatility proxy}
\begin{figure}[H]
\centering
\includegraphics[scale=0.45]{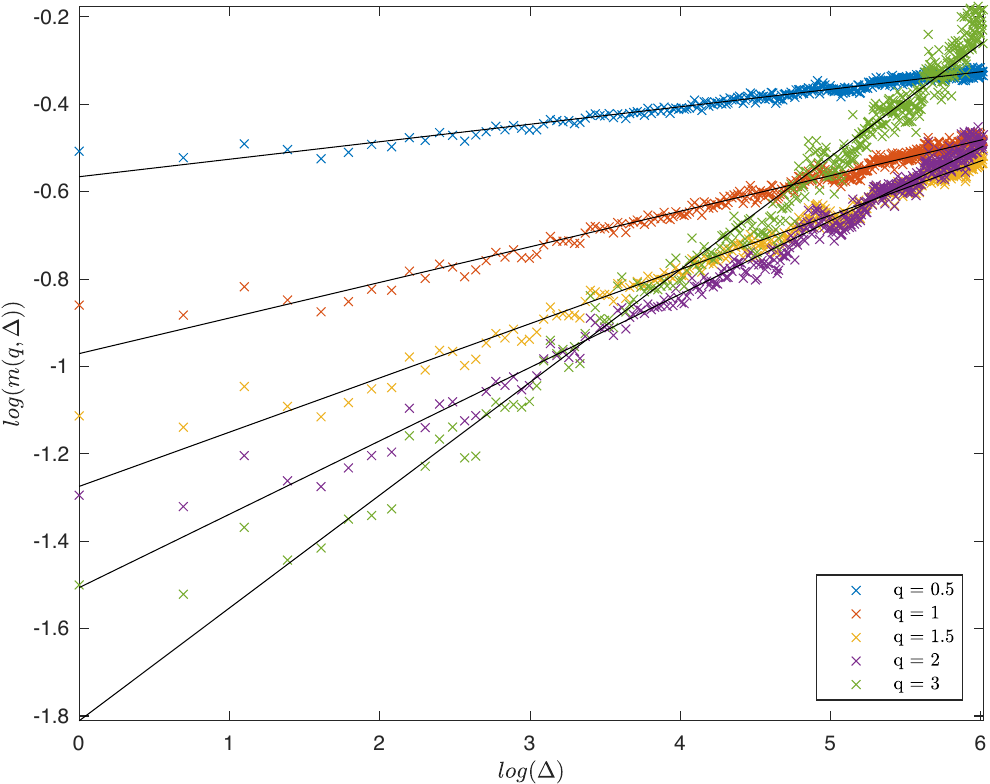} 
\includegraphics[scale=0.45]{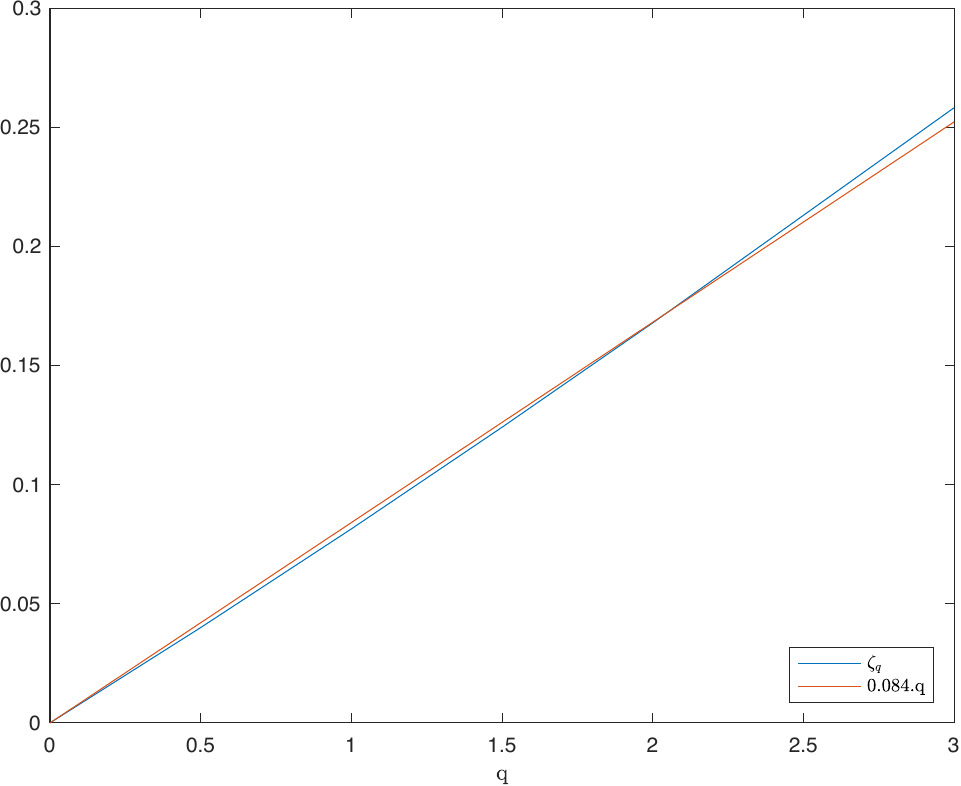} 
\caption{$\log m(q,\Delta)$ as a function of $\log \Delta$ (left), $\zeta_q$ (blue) and $0.082 \times q$ (green) (right), S\&P 100 (Parkinson volatility).}
\label{fig1SP100P}
\end{figure}

\begin{figure}[ht]
\centering
\includegraphics[scale=0.45]{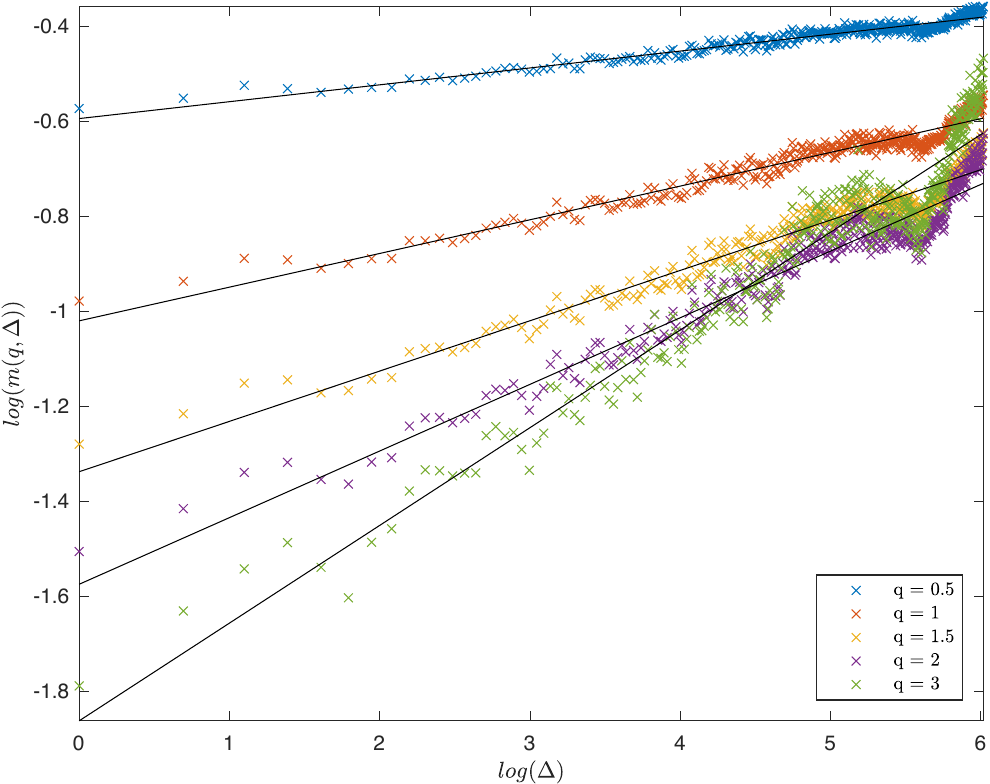} 
\includegraphics[scale=0.45]{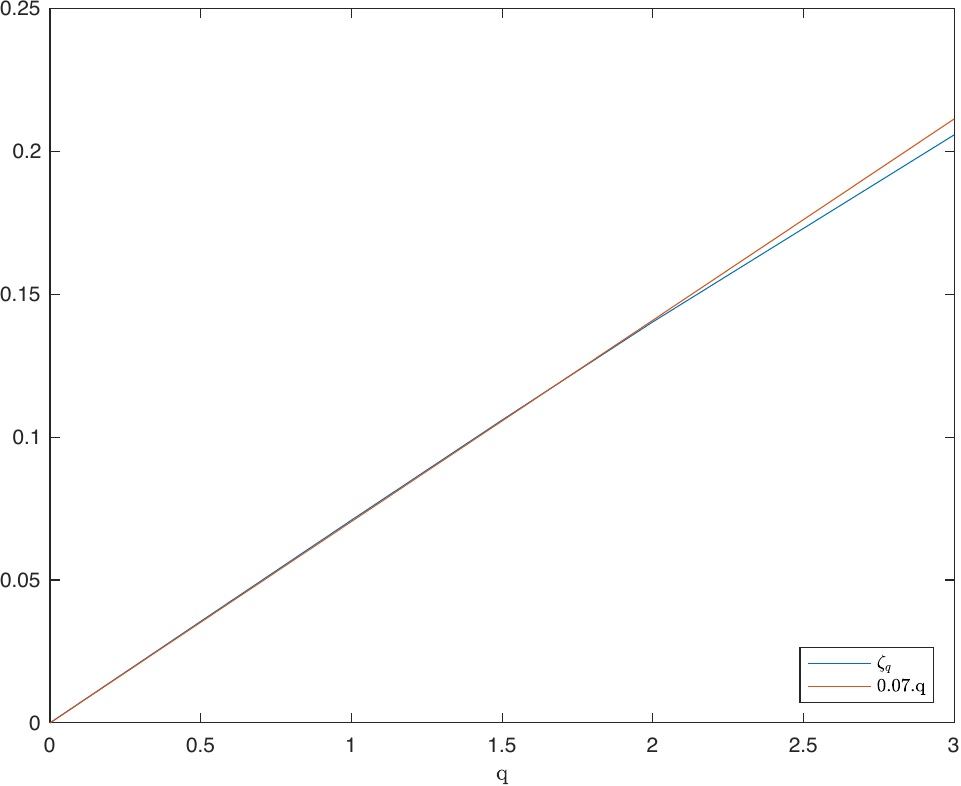} 
\caption{$\log m(q,\Delta)$ as a function of $\log \Delta$ (left), $\zeta_q$ (blue) and $0.064 \times q$ (green) (right), IBEX 35 (Parkinson volatility).}
\label{fig1IBEX35P}
\end{figure}

\noindent To ensure that the results apply to other range-based estimators, we reproduce the same analysis to the Parkinson volatility proxy based on the same data and on the same period. Results are expressed in Figures \ref{fig1SP100P} and \ref{fig1IBEX35P}.
\noindent The scaling is again linear in $q$ with $H = 0.082$ for the S\&P 100 and $H = 0.064$ for the IBEX 35. Values are very close to those found within the GK proxy. This confirms again that volatility based on the Parkinson proxy is rough.\\
\\
The smoothness parameter $H$ is detailed in Table \ref{Tab1P} for all the assets for the Parkinson volatility. One more time, we give the value for the whole period, split the period into two halves, and compute $H$ for each half.\\
\\
\begin{table}[ht]
\centering
\begin{tabular}{cccc}
Ticker & $H$ for the whole period & $H$ (first half) & $H$ (second half) \\
\hline
SP100 &0.0822 &0.0888 &0.0737\\
IBEX35 &0.0644 &0.0682 &0.0648\\
HSI &0.0452 &0.0555 &0.0336\\
MEXBOL &0.0638 &0.0738 &0.0489\\
FTSE100 &0.0774 &0.0823 &0.0669\\
ASX200 &0.0513 &0.0511 &0.0422\\
TOTAL &0.0738 &0.0856 &0.0608\\
XIN9I &0.0595 &0.0592 &0.0593\\
SHSZ300 &0.0591 &0.0668 &0.0499\\
BCOM &0.0127 &0.00623 &0.0251\\
INDU &0.08 &0.08 &0.0707\\
USDEUR &0.0265 &0.0276 &0.0261\\
IBOV &0.0694 &0.0746 &0.0603\\
MICROSOFT &0.0584 &0.0685 &0.0414\\
GOOGLE &0.0603 &0.063 &0.0542\\
SP400 &0.0757 &0.0822 &0.0623\\
\hline
\end{tabular}

\caption{Estimates of $H$ on the whole period and over two different time intervals for different indexes and stocks (Parkinson volatility)}
\label{Tab1P}
\end{table}
The same conclusions apply to Parkinson's volatility:
\begin{itemize}
\item $H$ remains between 0.01 and 0.09, with most assets around 0.07. This fact confirms that volatility is rough.	
\item $H$ is higher for the first period. This might be explained by the fact that this period (period 2005-2010) contains the 2008 crisis.
\item $H$ for Parkinson proxy of the BCOM asset is almost $0$.
\end{itemize}
\subsection{Distribution of the increments of the log-volatility}\label{subSect2}
\begin{figure}[H]
\centering
 \includegraphics[scale=0.9]{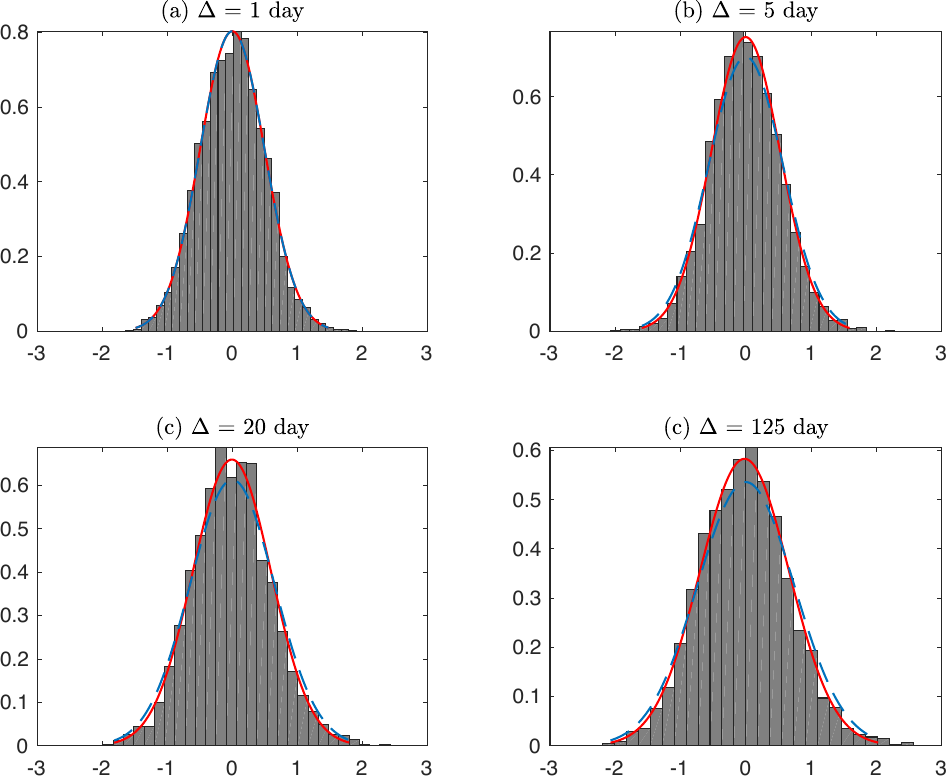} 
\caption{Histograms for various lags $\Delta$ of the increments $\log(\sigma_{t+\Delta)}) - \log(\sigma_t)$ of the Garman-Klass S\&P 100 log-volatility; the normal fit to distribution of the $\Delta$-days increments (red); normal fit to the 1-day increments rescaled by $\Delta^H$  (dashed blue)}
\label{fig3SP100}
\end{figure}

\noindent Now that we have established the common scaling behavior for our volatility proxies on the given data, we will focus on the S\&P 100 for the following results. Certainly, we will ensure that the same results are common for other assets, but unless specified otherwise, all the plots concern the S\&P 100.\\
\\
It is well-known that the distribution of log-volatility increments is very close to the normal distribution; see, for example, \cite{AB011}. We also find this in our data; see Figure \ref{fig3SP100}. Moreover, rescaling the density by $\Delta^H$ for any given lag $\Delta$ recovers the 1-day increments density. This is consistent with the fractional Brownian motion with Hurst parameter $H$, as seen in the previous section.\\
\section{RFSV model validation using range-based proxies}\label{Sect1.4}
\noindent In the following section, we test the accuracy of the Rough fractional volatility model (RFSV) introduced in \cite{GJR141} using the volatility data based on the Garman-Klass proxy. We will ensure that the model reproduces the same behavior as the data.
\subsection{The model}\label{SecModel}
\noindent Empirical results of Section \ref{Sect1.2} show that the increments of log-volatility based on range proxies for various assets appear to have a scaling property with constant Hurst parameter. We also made sure, through Section \ref{subSect2}, that their distribution is close to a normal distribution. Based on these results, the log-volatility increments can be modeled by the increments of a fBm as the following:
\begin{align}\label{EqSigma1}
\log \sigma_{t+\Delta} - \log \sigma_t = \nu \left(W_{t+\Delta}^H - W_t^H\right),
\end{align}
where $W^H$ is a fractional Brownian motion with the Hurst parameter estimated through the scaling of the volatility, and $\nu$ is a positive constant corresponding to the volatility of the increments. We can rewrite Equation (\ref{EqSigma1}) under the form:
\begin{align*}
\sigma_t = \sigma \exp(\nu W_t^H),
\end{align*}
where $\sigma$ is a positive constant.\\
\\
One of the drawbacks of this model is that it is not stationary. As a matter of fact, stationarity is a property that is desirable and useful for modeling time series. A possible model that keeps this property along with the fractional scaling is the fractional Ornstein-Uhlenbeck (fOU in short) process with a very long mean-reversion.\\
\\
The fractional Ornstein-Ublenbeck process $X_t$ is a stochastic process satisfying the stochastic differential equation:
\begin{align*}
dX_t =- \alpha X_t dt +  \nu dZ_t , \ \ X_0 = 0
\end{align*}
where both $\nu$ and $\alpha$ are positive parameters. When $Z_t$ is the standard Brownian motion, we get the standard Ornstein-Uhlenbeck, see \cite{OU301}. Our interest, however, is the when $Z_t = W_t^H$. We also consider an arbitrary initial point $X_0 = m$ instead of $0$. The SDE followed by the the process of log-volatility becomes:
\begin{align}\label{EqfOU1}
dX_t =- \alpha(X_t - m)dt +  \nu dW_t^H, \ \ x_0 = m
\end{align}
where $m \in \mathbb{R}$ and $(W_t^H)$ is the fBm with Hurst parameter $H$.\\
\\
Equation (\ref{EqfOU1}) is then solved using the following explicit representation:
\begin{align}\label{EqfOU2}
X_t = \nu \int_{-\infty}^{t}{e^{-\alpha(t-s)}dW_t^H + m}
\end{align}
where the stochastic integral with respect to fBM is simply a path-wise Riemann-Stieljes integral (\cite{CKM031}). Lastly, we recover the volatility and thus define the RFSV model on the time interval [0,T]:
\begin{align*}
\sigma_t = \exp\left(X_t\right), \ \ t \in [0,T],
\end{align*}
where $(X_t)$ satisfies equation (\ref{EqfOU2}) for some $\nu > 0$, $\alpha > 0$, $m \in \mathbb{R}$ and $H < 1/2$ the measured smoothness of the volatility. In addition to the stationarity of such a model, choosing $\alpha \ll 1/T$ allows the log-volatility to behave locally (at time scales smaller than $T$) as a fBm. This observation is formalized by Proposition 3.1 in \cite{GJR141} we recall below:
\begin{proposition}\label{Prop1}
Let $W^H$ be a fBm and $X^\alpha$ defined by (\ref{EqfOU2}) for a given $\alpha > 0$. As $\alpha$ tends to zero,
\begin{align*}
\mathbb{E}[\sup_{t\in [0,T]}\mid X_t^\alpha - X_0^\alpha - \nu W_t^H \mid] \rightarrow 0
\end{align*}
\end{proposition}
\paragraph{Remarks:}
\begin{itemize}
\item Proposition (\ref{Prop1}) implies that within the interval [0,T], and under the condition $\alpha \ll 1/T$, we can proceed as if the log-volatility process were a fBm. Setting $\alpha = 0$ allows to recover the simple non-stationary fBm (\ref{EqSigma1}).
\item The RFSV differs from the classical FSV model of Comte and Renault, see \cite{CR981}, in that, instead of taking $H > 1/2$ and $\alpha$ large in FSVr, the RFSV model is defined for $H < 1/2$ and $\alpha$ small (actually $\alpha$ is chosen not to be equal to $0$ only so that the volatility satisfies stationarity). 
\item The choice $H < 1/2$ is consistent with both the statistical properties of the data and generates a term structure of volatility skew that matches the observations.
\item The choice of the fOU is for convenience and simplicity. Other models that imitate the fBm behavior at reasonable time scales and are stationary can be considered as well. 
\item The RFSV process reproduces approximately the exact scaling property as the fBm. This is a consequence of the following corollary:
\begin{corollary}\label{coro1}
Let $q>0$, $t>0$, $\Delta >0$. As $\alpha$ tends to zero, we have:
\begin{align*}
\mathbb{E}[\mid X_{t+\Delta}^\alpha - X_t^\alpha \mid^q] \rightarrow \nu^q K_q \Delta^{qH}.
\end{align*}
\end{corollary}
A detailed comparison between the RFSV and FSV can be found in \cite{GJR141}.
\end{itemize}
\subsection{Model validation}
\begin{figure}[ht]
	\centering
\includegraphics[height = 7cm,width = 14cm]{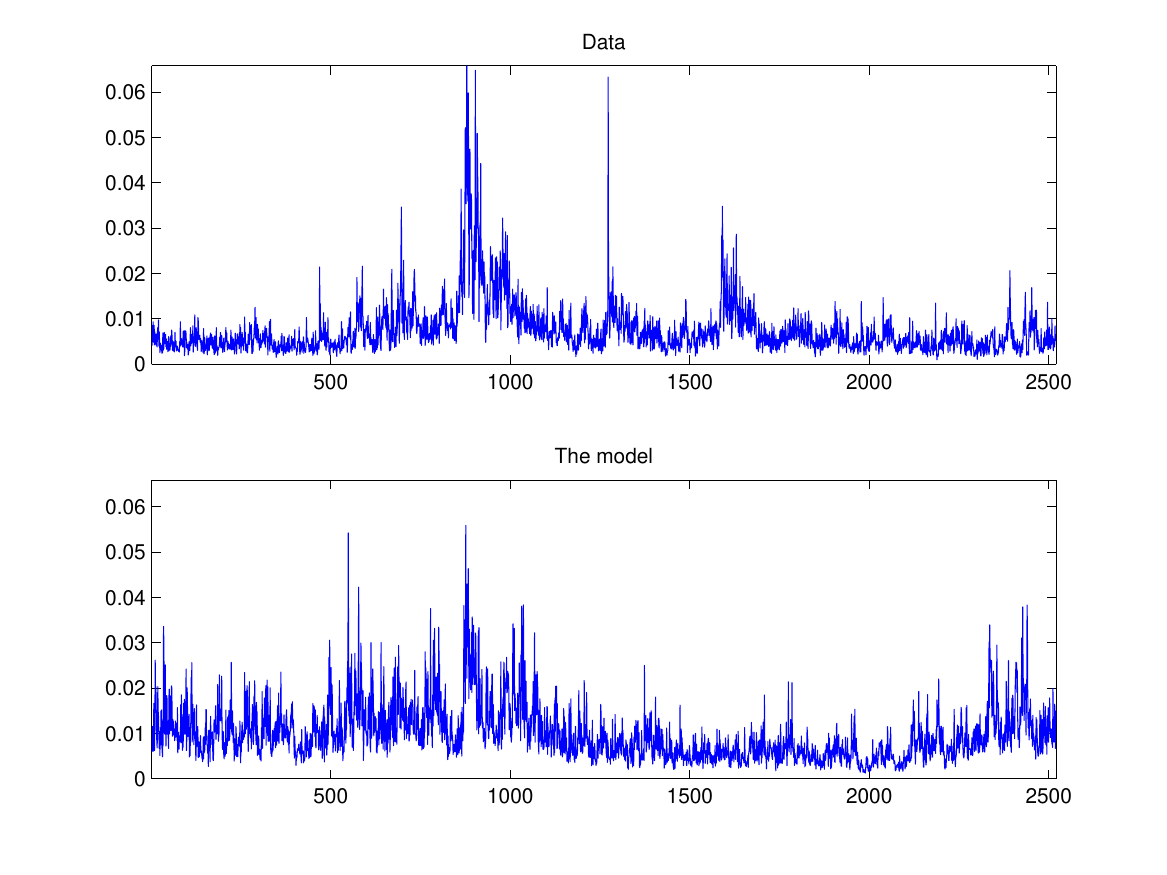}
\caption{Garman Klass volatility of S\&P 100 (above) and simulated paths (below)}
\label{figDataVsModel}
\end{figure}

\begin{figure}[H]
	\centering
\includegraphics[scale=0.48]{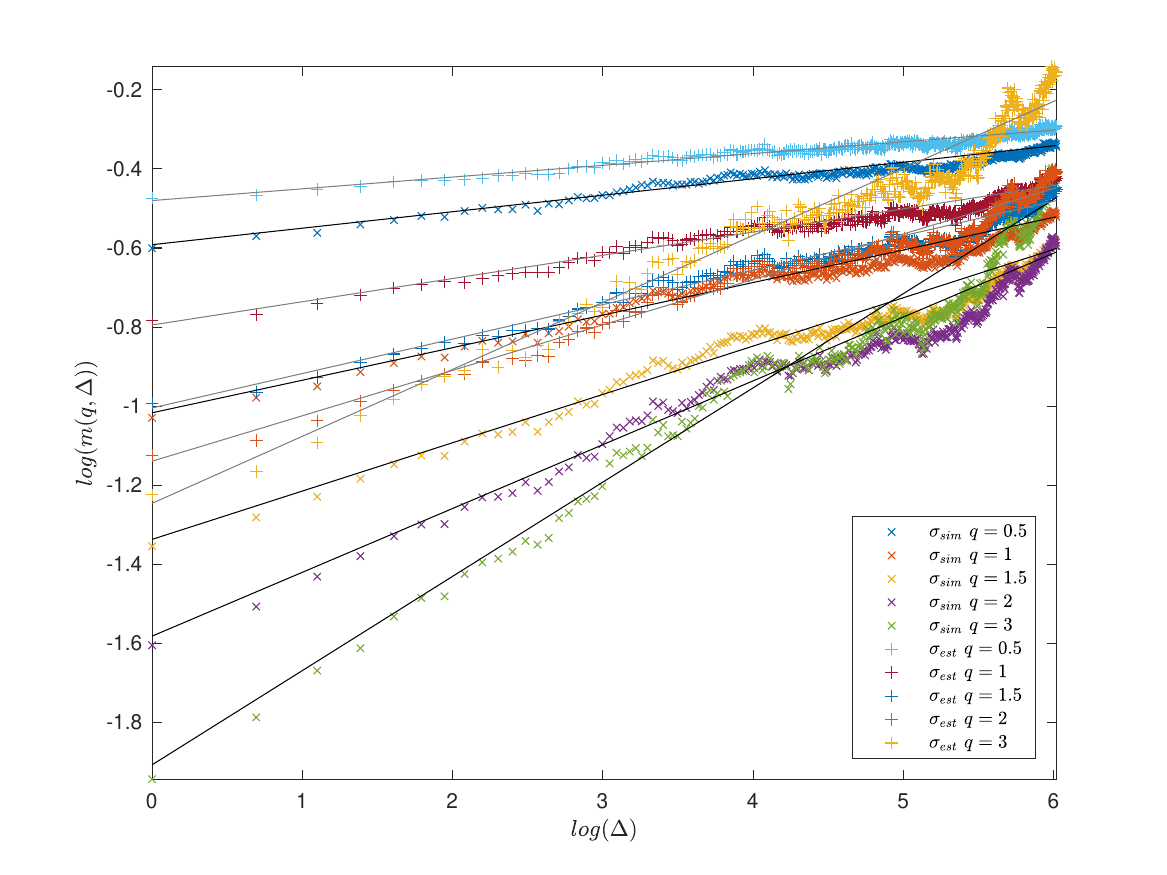} 
\caption{$\log(m(q,\Delta))$ as a function of $\log(\Delta)$ for the real spot volatility with $H = 0.08$ and the Garman Klass proxy based on simulations.}
\label{figDataVsModelSmooth}
\end{figure}

\noindent In previous sections, we come to the conclusion that the RFSV model seems to be a relevant volatility model based on empirical results. But the question that arises is whether the estimated range-based volatility proxies, from simulated data with RFSV as the volatility process, behave like the underlying process (in terms of the scaling properties). Our goal is to investigate this question. To do so, we simulate the spot volatility process using the RFSV model, simulate intraday prices, recover range prices (open, close, high, and low), and finally estimate the range volatility from the simulated range prices. On the one hand, we will be able to compare the behavior of the known real spot volatility process to that of the proxy used for its approximation. We will also be able to estimate the realized volatility and compare it to the Garman-Klass and to real data.\\
\\
Spot volatility is simulated using the RFSV model for $2,521$ days. Since range-based volatility assumes the volatility process to be constant within the day, we take into account the randomness of the intraday prices $P_u$ where $u$ frequency is of the order of a few seconds and simulate the spot volatility $\sigma_t$ on a daily basis. We choose parameters that are consistent with the S\&P 100 Garman-Klass volatility (even though a precise estimation for $\nu$ $\alpha$ and $m$ is not simple) and consistent with our empirical estimates from Section 1.3, i.e., $H = 0.08$, $\nu = 0.3$, $m = X_0 = -5$ and $\alpha = 5\times 10^{-4}$.\\
\\
To simulate the volatility and price paths, we proceed as the following :
\begin{itemize}
\item Simulate fBm using a wavelet-based synthesis, see  \cite{AS961}.
\item Simulate the log-volatility process $X$ for each day $n$ according to a discrete scheme:
\begin{align*}
 X_{n+1} - X_{n} = \nu\left(W_{n+1}^H - W_{n}^H \right) + \alpha (m-X_{n}).
\end{align*}
\item Simulate the asset price $P$ by taking:
\begin{align*}
P_{n + (j+1)\delta} - P_{n +j\delta} = P_{n + j\delta}\sigma_{n}\sqrt{\delta}U_{j},
\end{align*}
where the $U_j$ are iid standard Gaussian variables.
\item Extract range prices  (open, close, high, and low prices) for each day and compute the realized volatility and the Garman-Klass range-volatility.
\end{itemize}
We present in Figure \ref{figDataVsModel} a plot of the Garman-Klass proxy from S\&P 100 data along with the simulated spot volatility described above.\\
\\
\noindent We compare the plots of the S\&P 100 Garman-Klass proxy with the simulated path. Graphically, it seems that estimated volatility exhibits the same behavior as the simulation, at least to a visual extent. This was already verified in \cite{GJR141}. A zoom-in or a zoom-out gives typically the same kind of qualitative properties.
To compare the smoothness of the real spot volatility (simulated paths), with that of the Garman-Klass proxy recovered from simulated prices, we repeat the analysis of Section \ref{Sect1.2}. We plot in Figure \ref{figDataVsModelSmooth} $\log(m(q,\Delta))$ as a function of $\log(\Delta)$ for both the real volatility and volatility proxy. The Hurst exponent of the estimated Garman-Klass proxy is relatively close to the true one ($H = 0.079$ compared to $H = 0.056$). The mismatch can in particular, be due to simulation bias.
\subsection{FSV vs. RFSV}
\begin{figure}[ht]
\centering
\includegraphics[height = 5cm, width = 15cm]{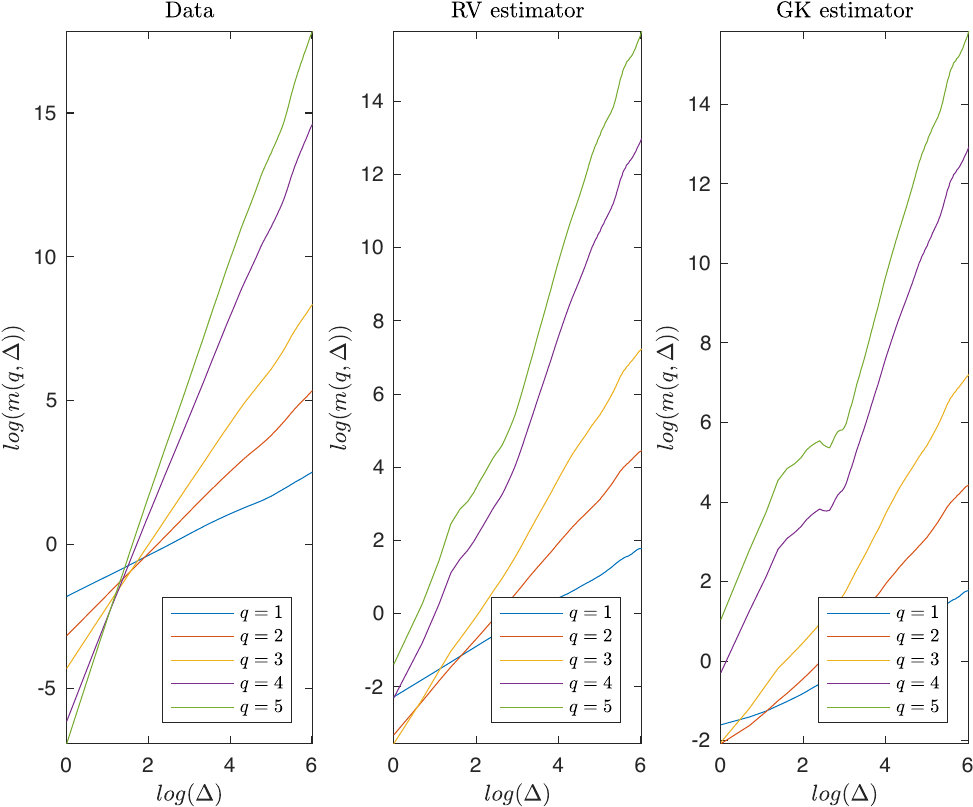}
\caption{$\log(m(q,\Delta))$ as a function of $\log(\Delta)$ for the data (left) Realized Volatility (center) and Garman-Klass volatility (right) over the simulated paths for $q = 1,2,3,4,5$}
\label{fig:FSVH07}
\end{figure}

\begin{figure}[ht]
\centering
\includegraphics[width=14cm,height=5.5cm]{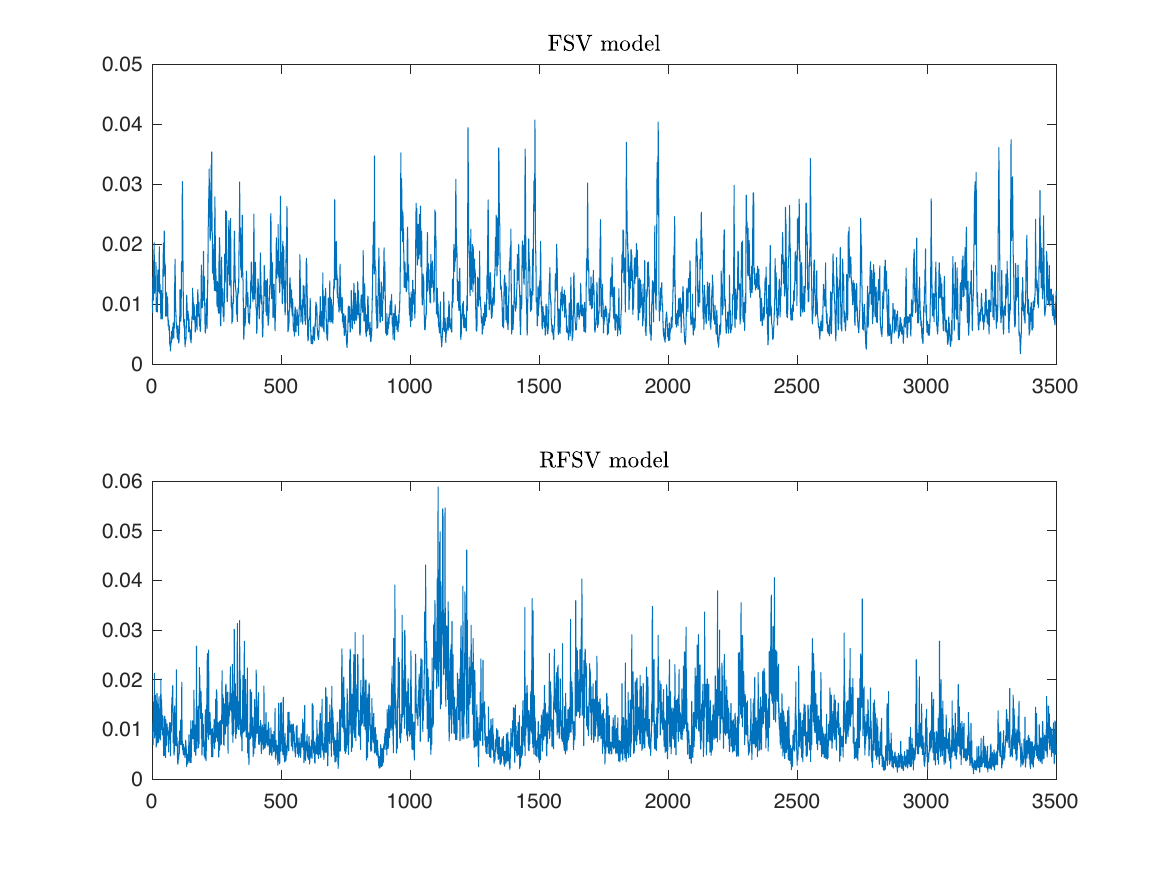}\\
\includegraphics[width=14cm,height=5.5cm]{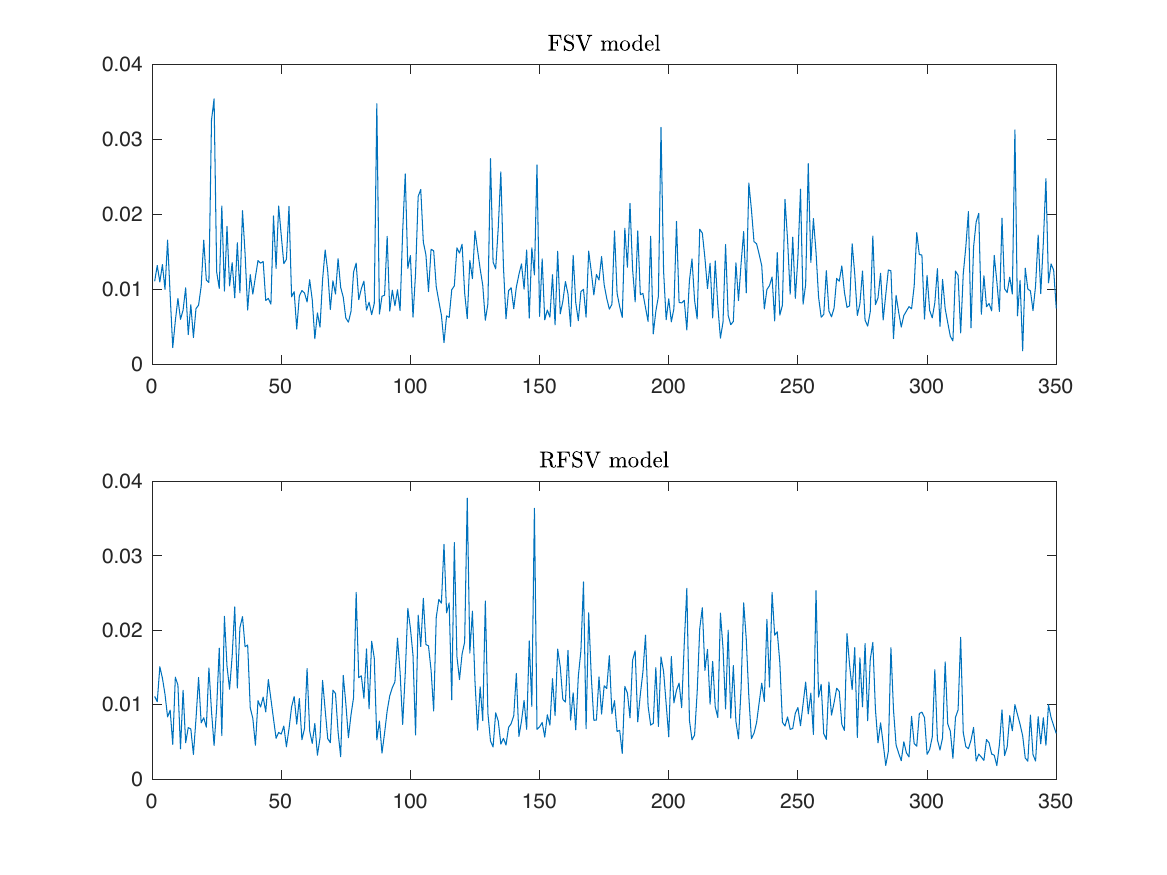}
\caption{Plot of one path of the FSV and RFSV models for $H = 0.7$ and $H = 0.08$ observed daily (top two), and every 10 days (bottom two).}
\label{fig:FSVvsRFSV}
\end{figure}

\begin{figure}[ht]
\centering
\includegraphics[width=15cm,height=6cm]{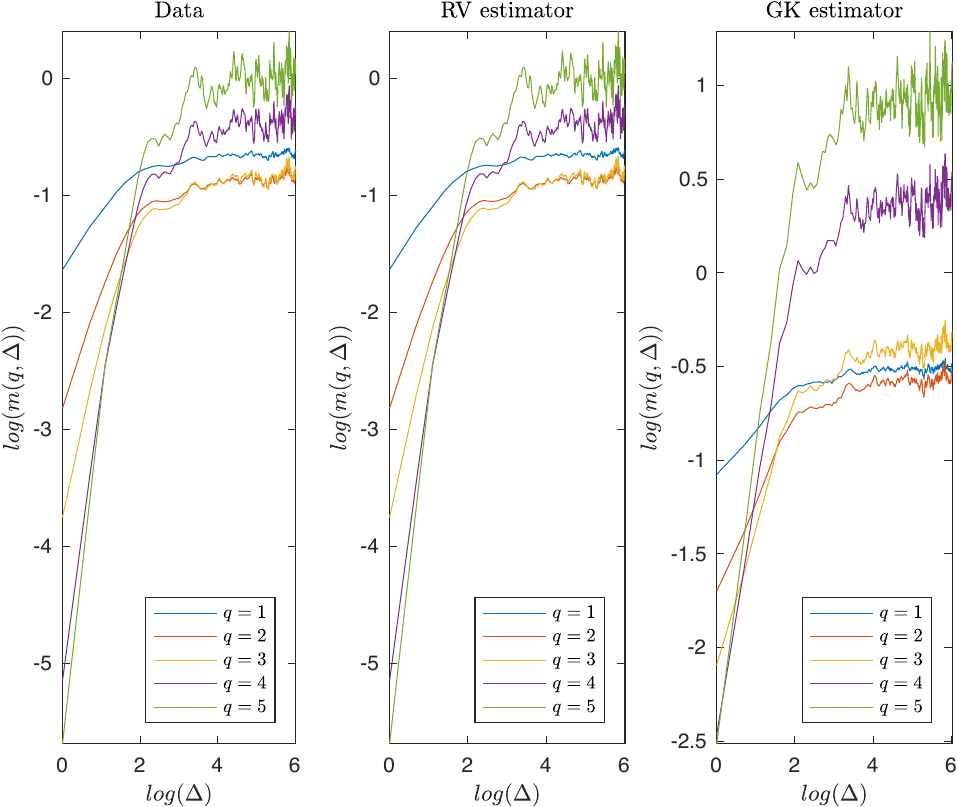}\\
\includegraphics[width=15cm,height=6cm]{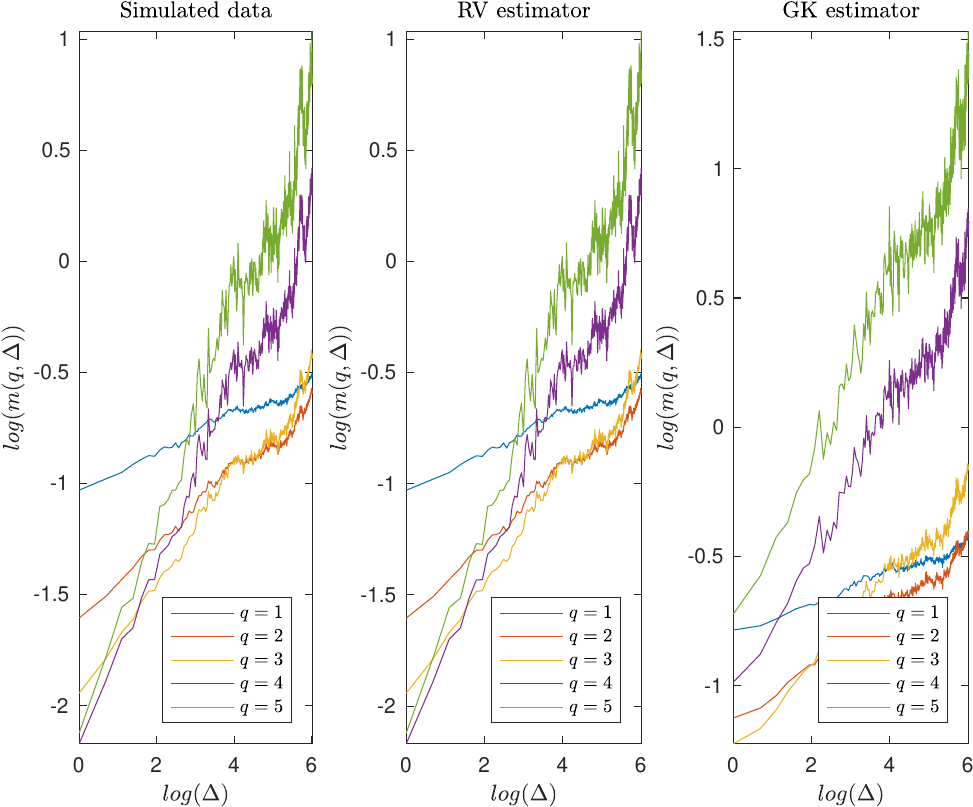}
\caption{$\log(m(q,\Delta))$ as a function of $\log(\Delta)$ for the simulated paths (left), Realized Volatility (center), and Garman Klass volatility (right) over the simulated paths for $q = 1,2,3,4,5$, for FSV model (top) and RFSV model (bottom)}
\label{fig:FSVvsRFSVSmoothing0}
\end{figure}

\begin{figure}[ht]
\centering
\includegraphics[width=15cm,height=6cm]{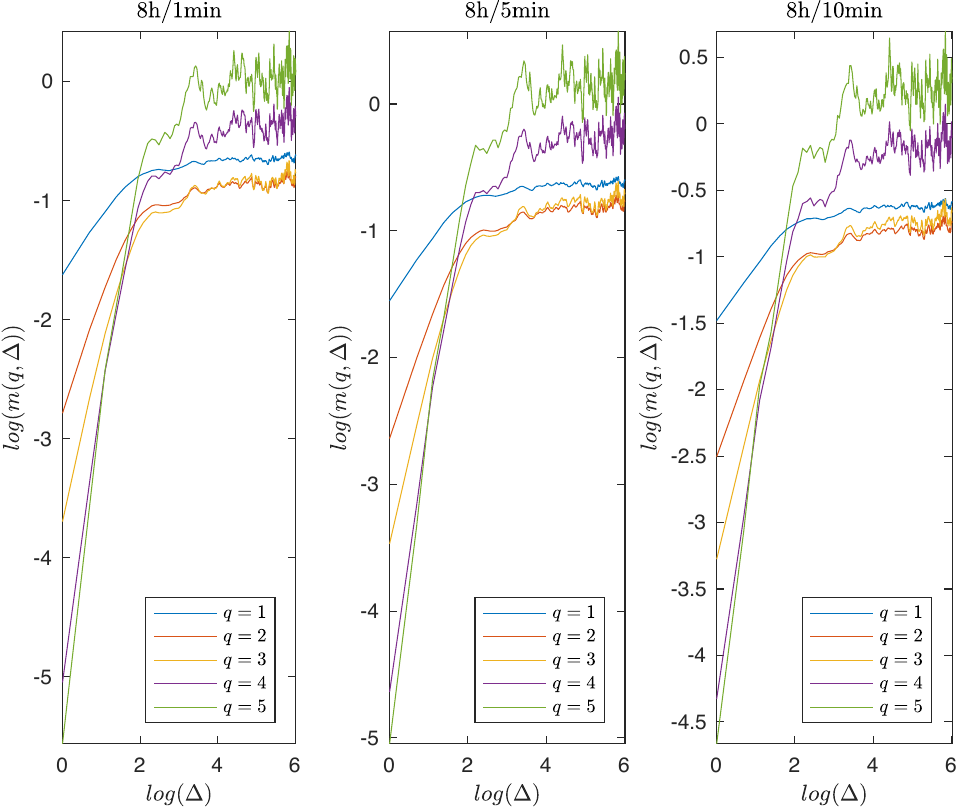}\\
\includegraphics[width=15cm,height=6cm]{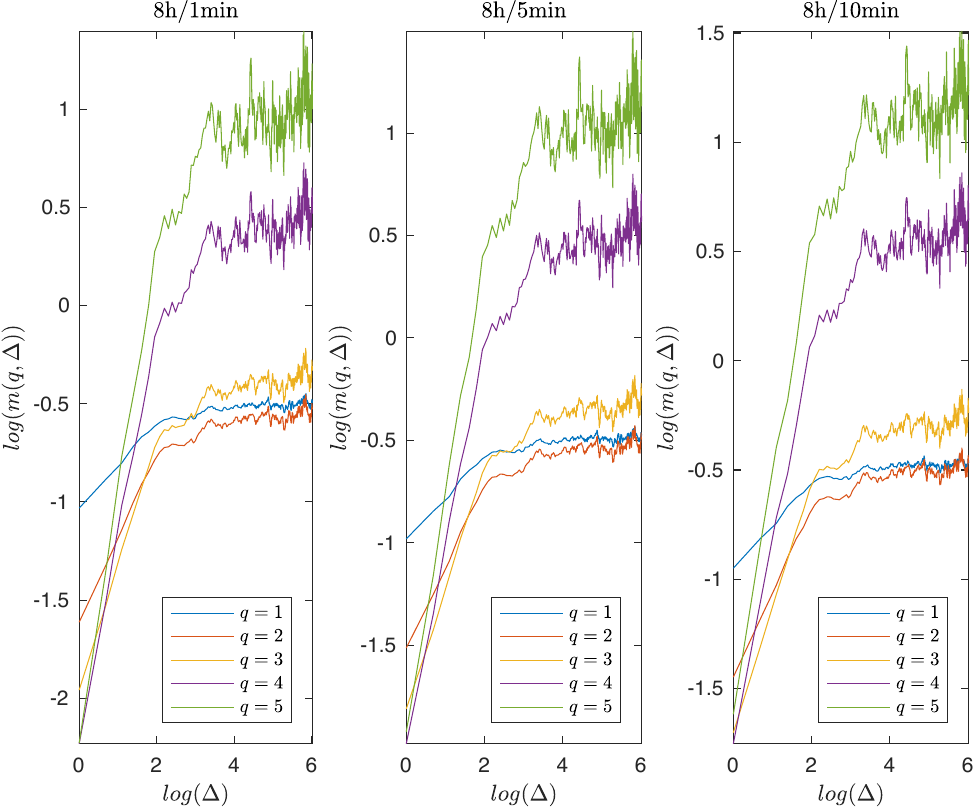}
\caption{$\log(m(q,\Delta))$ as a function of $\log(\Delta)$ for realized volatility (top) and Garman-Klass volatility (bottom) based on an 8 hours window with 1min (right), 5min (center) and 10min (left) discretization for the mean-reverting FSV model with $H = 0.07$.}
\label{fig:FSVvsRFSVSmoothing1}
\end{figure}

\noindent We have shown from our analysis that the RFSV model hypothesis can not be rejected. In light of empirical results on data, we would like to test if we can reject the FSV model by simulating the volatility process. To do so, we analyze the behavior of the smoothing function $m(q,\Delta)$ for small and large lags, using simulated volatility and asset prices, and see if the Garman-Klass and realized volatility based on these simulations behave like the one found for real data.\\
\\
\noindent The first basic difference between RFSV and FSV is the range of the Hurst exponent values for the fractional Brownian motion. First, we consider the simple non-mean-reverting fractional volatility model with $H > 0.5$, i.e. $\sigma_t = \sigma_0e^{\nu W_t^H}$ with $H = 0.7$ and $\eta = 0.25$. We take a look at the scaling behavior of the realized-volatility and Garman-Klass volatility proxies based on this model in Figure \ref{fig:FSVH07}.
\noindent In Figure \ref{fig:FSVH07}, we see that the scaling of $\log(\Delta)$ is very close to a straight line for the RV and GK estimators. The resulting smoothing parameter found in these figures is close to the original one ($H = 0.69$ for the RV estimator and $H = 0.64$ for the GK estimator).\\
\\
We are aware that such a model, leading to crazy volatility values, does not make sense without mean-reversion. However, this allows us to exclude a fractional volatility model of the form $\sigma_t = \sigma_0e^{\nu W_t^H}$ with $H > 0.5$.\\
\\
\noindent Unlike the RSFV model where the mean-reversion is intrinsic to the model for $\alpha = 0$, Comte and Renault impose $\alpha$ to be large enough, i.e. $\alpha \gg 1/T$ where $T$ is the time horizon of interest, to verify this property. We would like to test to what extent this model can be misleading in estimating the smoothness of the diffusion process. We compare the FSV and RFSV models for the set of parameters given in Table \ref{tab:FSVvsRFSV}, which leads to the simulated time series given in Figure \ref{fig:FSVvsRFSV}.
\begin{table}[ht]
\centering
\begin{tabular}{lll}
\hline 
  & FSV & RFSV \\ 
\hline 
$H$ & $0.7$ & $0.08$ \\
$\alpha$ & $0.25$ & $5\times 10^{-4}$  \\ 
$\nu$ & $0.25$ & $0.45$  \\
$m$  & $-4.5$ & $-5$ \\
$X_0$ & $-4.5$ & $-5$ \\
$\mathbb{E}[\log(\sigma)]$ & $-4.6$  & $-4.7$  \\
$\mathbb{V}ar[\log(\sigma)]$ & 0.21 & 0.33  \\
\hline 
\end{tabular}
\caption{Parameters values used for simulating the FSV and RFSV models and the mean and variance of the simulated time series}
\label{tab:FSVvsRFSV}
\end{table}
\noindent Figure \ref{fig:FSVvsRFSV} shows that both processes satisfy mean-reversion. The roughness of the RFSV model is quite clear when we observe the process daily, however, when observed every 10 days, it is less obvious to say which is the FSV and which is the RFSV. \\
\\
Figure \ref{fig:FSVvsRFSVSmoothing0} checks one more time the scaling behavior of the two processes and their RV and GK proxies estimated on 24-hour windows and 1-second observed prices.
\noindent We confirm through Figure \ref{fig:FSVvsRFSVSmoothing0} the following key results:
\begin{itemize}
	\item The smoothing function $\zeta_q$ for the RFSV model keeps the same pattern for short and long time scales, ($\log(m(q,\Delta))$ is close almost linear w.r.t $\log(\Delta)$).
	\item The FSV model seems to exhibit two slopes. At small scales, the slope is close to that of the Hurst exponent of the fractional Brownian motion that drives the process, i.e. $H \approx 0.7$. At large scales, the slope gives a value close to $0$. Actually, the stationarity of the process at large scales is responsible for such estimation.\\
\end{itemize} 
\noindent Finally, in order to verify the impact of discretization on the estimators, we consider that prices are observed on an 8-hour time window every 1, 5, or 10 minutes. We compute the RV and GK estimators based on these observations. Results are shown in Figure \ref{fig:FSVvsRFSVSmoothing1} for the FSV (this test on the RFSV model does not bring more information).
\noindent We observe an other phenomenon for the GK estimator. The slope on very small scales ($\Delta = 1,2$) gives a relatively smaller smoothing parameter ($H$ around $0.19$) than for intermediate lags ($H$ around $0.24$). We believe that this is due to the noise of this estimator. Moreover, the value of the smoothing parameter is still very small compared to the one used to simulate the process.\\
\\
Finally, we can conclude that FSV volatility with $H > 0.5$ behaves differently than the data. It may therefore be excluded from being a good volatility model. Rough fractional model, on the other hand, seems at this point, the most plausible to model the volatility. Even though it was not illustrated here, mean-reversion models with standard Brownian diffusion are even less plausible.
\subsubsection*{Quantitative justification}
\noindent In this section, we would like to quantify the phenomena encountered previously. Our goal is to see how estimating the smoothing parameter is affected by the lag range. We denote by $X_t$ the asset log price $X_t = \log(P_t)$. Since spot volatility does not really make sense at the intraday level (beyond seasonality), and to avoid the smoothing issue of realized volatility estimation, we assume that the volatility is constant within a day, i.e., for a day $i$ and for each time $t \in [i,i+1)$ $\sigma_t = \sigma_i = \text{ constant }$, with $\sigma_i = e^{\nu W_i^H}$ (although this model does not make much sense for $H > 0.5$).\\
\\
Assume we have $n$ observations for each day $i$ for the price process $P_i^{j = 1,...,n}$, and let $\Delta_j^n X$ the log price increments of day $i$:
\begin{align*}
\Delta_j^n X = \log(P_{i}^{j}) - \log(P_{i}^{j-1}).
\end{align*}
Using the central limit theorem (CLT) for realized volatility we have:
\begin{align*}
(\sigma_i^{RV})^2 = & \sum_{j = 1}^{n}(\Delta_j^n X)^2 \\
\simeq & \int_i^{i+1}e^{2\nu W_i^H}dt + \frac{1}{\sqrt{n}}\sqrt{2\int_i^{i+1}e^{4\nu W_i^H}dt}\xi \text{ where } \xi \sim \mathcal{N}(0,1) \\
\simeq & e^{2\nu W_i^H}\big(1 + \sqrt{\frac{2}{n}}\xi\big),
\end{align*}
which leads to the approximation:
\begin{align*}
\log(\sigma_i^{RV}) \simeq \nu W_i^H + \sqrt{\frac{1}{2n}}\xi \text{ where } \xi \sim \mathcal{N}(0,1).
\end{align*}
Taking the increments of the log volatility between $i$ and $i + \Delta$ for a given time lag $\Delta$, we have:
\begin{align*}
\log(\sigma_{i+\Delta}^{RV}) - \log(\sigma_{i}^{RV})  \simeq \underbrace{\nu\big(W^H_{i+\Delta} - W^H_{i}\big)}_{O\big(\nu\Delta^H\big)} + \underbrace{\sqrt{\frac{1}{2n}}(\xi + \xi' )}_{O\big(\sqrt{\frac{1}{n}}\big)},
\end{align*}
where $\xi$ and $\xi'$ are i.i.d Gaussian variables.\\
\\
\noindent This equivalence leads to the following observations:
\begin{itemize}
	\item When the vol of vol $\nu$ is small,i.e. such $\nu^2\Delta^{2H} \ll \frac{1}{n}$, the noise takes the upper hand. As a result, we would observe a slope close to horizontal when plotting $\log(m(q,\Delta))$ against $\log(\Delta)$.
	\item On the contrary, when $\nu$ is large, the first term is predominant and the slope is close to $qH$ which allows for a rather more precise estimation of the true Hurst exponent.
	\item In the intermediate case, two slopes can be observed; one corresponds to estimation noise, and the other is proportional to the true Hurst exponent. 
\end{itemize}
\noindent We illustrate these cases in Figure \ref{figquantjustif} below.
\begin{figure}[H]
	\centering
\includegraphics[width=5cm,height=4cm]{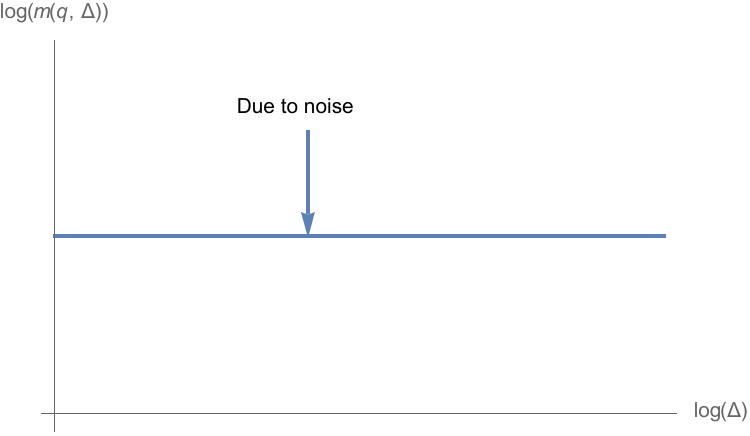}
\includegraphics[width=5cm,height=4cm]{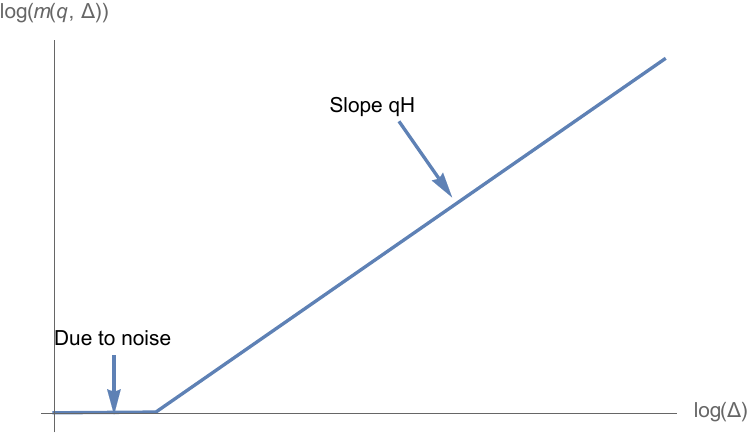}
\includegraphics[width=5cm,height=4cm]{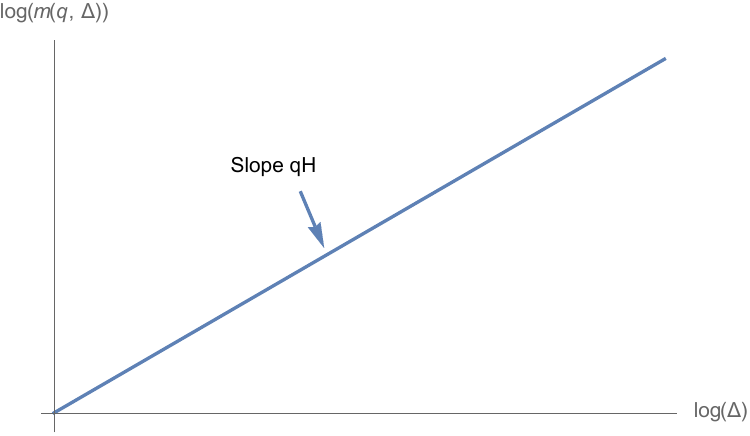}
\caption{Different cases encountered for the estimation of the slope of $\log(m(q,\Delta))$ against $\log(\Delta)$.}
\label{figquantjustif}
\end{figure}
\section{Forecasting range-based volatility using the RFSV model}\label{Sect1.5}
\noindent We conclude the paper by assessing the prediction accuracy of of the RFSV model compared to other common models such as the AR, HAR, or GARCH. 
\subsection{Forecasting log-volatility}\label{SectionForecastLogVol}
\noindent The key formula on which the prediction method is based is the following one:
\begin{align*}
\mathbb{E}[W_{t+\Delta}^H \mid \mathcal{F}_t] = \frac{\cos(H\pi)}{\pi}\Delta^{H+1/2}\int_{-\infty}^{t}{\frac{W_s^H}{(t-s+\Delta)(t-s)^{H+1/2}}}ds,
\end{align*}
where $W^H$ is a fBm with $H < 1/2$ and $\mathcal{F}_t$ the filtration it generates, see Theorem 4.2 of \cite{NP001}. By construction, over any reasonable time scale of interest, as formalized in Corollary \ref{coro1}, we may approximate the fOU volatility process in the RFSV model as $\log \sigma_t^2 \approx 2\nu W_t^H + C$ for some constants $\nu$ and $C$. Our prediction formula for the log-variance then follows:
\begin{align}\label{predictLogSigmaSquare}
\mathbb{E}[\log \sigma_{t + \Delta}^2 \mid \mathcal{F}_t] = \frac{\cos(H\pi)}{\pi}\Delta^{H+1/2}\int_{-\infty}^{t}{\frac{\log \sigma_s^2}{(t-s+\Delta)(t-s)^{H+1/2}}}ds.
\end{align}
This formula, or rather its approximation through a Riemann sum (we assume in this section that volatilities are perfectly observed, although they are in fact estimated), is used to forecast the log-volatility 1,5 and 20 days ahead ($\Delta = 1, 5, 20$).\\
\\
In the spirit of \cite{Cor091}, we compare the predictive power of Formula (\ref{predictLogSigmaSquare}) with that of AR, HAR, and GARCH forecasts. Recall that for a given integer $p > 0$, the AR(p) and HAR predictors take the following form (where the index $i$ runs over the series of daily volatility estimates):
\begin{itemize}
\item AR(p):
\begin{align*}
\widehat{\log(\sigma_{t+\Delta}^2)} = K_0^\Delta + \sum_{i = 0}^{p}C_i^\Delta \log(\sigma_{t-i}^2).
\end{align*}
\item HAR:
\begin{align*}
\widehat{\log(\sigma_{t+\Delta}^2)} = K_0^\Delta + C_0^\Delta \log(\sigma_t^2) + C_5^\Delta\frac{1}{5}\sum_{i = 0}^{p}C_i^\Delta \log(\sigma_{t-i}^2) + C_{20}^\Delta\frac{1}{20}\sum_{i=0}^{20}\log(\sigma_{t-i}^2).
\end{align*}
\end{itemize}
We estimate AR and HAR using a rolling time window of 500 days. For the HAR case, we use standard linear regression to estimate the coefficients as explained in \cite{Cor091}. In the sequel, we consider $p = 5$ and $p=10$ in the AR formula. Indeed, these parameters essentially give the best results for the horizons at which we wish to forecast the volatility (1, 5, and 20 days). For each day, we forecast volatility for five different indexes.\\
\\
We then assess the quality of the various forecasts by computing the ratio $P$ between the mean squared error of our predictor and the approximated variance of the log-variance:
\begin{align*}
P = \frac{\sum_{k = 500}^{N-\Delta}{\left(\log(\sigma^2_{k+\Delta}) - \widehat{\log(\sigma_{t+\Delta}^2)} \right)^2}}{\sum_{k = 500}^{N-\Delta}{\left(\log(\sigma^2_{k+\Delta}) - \mathbb{E}[\log(\sigma^2_{k+\Delta})] \right)^2}},
\end{align*}
where $\mathbb{E}[\log(\sigma^2_{k+\Delta})]$ denotes the empirical mean of the log-variance over the whole period.\\
\\
We present in Table \ref{Tab1} the ratio $P$ for different models in order to compare the RFSV prediction power with other autoregressive models for predicting the log variance.
\begin{center}\label{Tab1.2}
\begin{longtable}{| p{.30\textwidth} | p{.10\textwidth} || p{.10\textwidth} | | p{.10\textwidth} | | p{.10\textwidth} |} 
\caption{Ratio P for AR, HAR and RFSV predictors for $\log(\sigma_{t+\Delta}^2)$}\\
\hline
Ticker  & AR(5) & AR(10) & HAR(3) & RFSV \\
\hline
SP100 $\Delta = 1$ &{0.451} &0.446 &\textbf{0.443} &0.466\\
SP100 $\Delta = 5$ &0.644 &0.635 &\textbf{0.546} &0.557\\
SP100 $\Delta = 21$ &0.897 &0.894 &0.734 &\textbf{0.718}\\
\hline 
IBEX35 $\Delta = 1$ &0.594 &0.594 &\textbf{0.582} &0.622\\
IBEX35 $\Delta = 5$ &0.843 &0.824 &\textbf{0.728} &\textbf{0.728}\\
IBEX35 $\Delta = 21$ &1.18 &1.17 &0.943 &\textbf{0.908}\\
\hline 
HSI $\Delta = 1$ &0.529 &0.523 &\textbf{0.513} &0.52\\
HSI $\Delta = 5$ &0.647 &0.633 &\textbf{0.575} &0.577\\
HSI $\Delta = 21$ &0.805 &0.801 &\textbf{0.665} &0.671\\
\hline 
MEXBOL $\Delta = 1$ &0.572 &0.567 &\textbf{0.553} &0.589\\
MEXBOL $\Delta = 5$ &0.731 &0.709 &0.648 &\textbf{0.645}\\
MEXBOL $\Delta = 21$ &0.922 &0.917 &\textbf{0.757} &0.764\\
\hline 
FTSE100 $\Delta = 1$ &0.474 &0.465 &\textbf{0.463} &0.476\\
FTSE100 $\Delta = 5$ &0.627 &0.614 &\textbf{0.545} &\textbf{0.545}\\
FTSE100 $\Delta = 21$ &0.859 &0.855 &0.699 &\textbf{0.688}\\
\hline 
ASX200 $\Delta = 1$ &0.536 &\textbf{0.52}4 &\textbf{0.524} &0.527\\
ASX200 $\Delta = 5$ &0.658 &0.652 &0.577 &\textbf{0.573}\\
ASX200 $\Delta = 21$ &0.806 &0.793 &0.707 &\textbf{0.688}\\
\hline 
TOTAL $\Delta = 1$ &0.540 &0.534 &\textbf{0.527} &0.558\\
TOTAL $\Delta = 5$ &0.720 &0.704 &0.640 &\textbf{0.636}\\
TOTAL $\Delta = 21$ &1.008 &1.015 &0.809 &\textbf{0.789}\\
\hline 
XIN9I $\Delta = 1$ &0.587 &0.58 &\textbf{0.568} &0.582\\
XIN9I $\Delta = 5$ &0.712 &0.695 &\textbf{0.637} &0.641\\
XIN9I $\Delta = 21$ &0.913 &0.918 &0.762 &\textbf{0.758}\\
\hline 
SHSZ300 $\Delta = 1$ &0.574 &0.568 &\textbf{0.56} &0.572\\
SHSZ300 $\Delta = 5$ &0.707 &0.695 &\textbf{0.634} &\textbf{0.634}\\
SHSZ300 $\Delta = 21$ &0.896 &0.904 &0.772 &\textbf{0.753}\\
\hline 
BCOM $\Delta = 1$ &0.846 &0.838 &\textbf{0.805} &0.83\\
BCOM $\Delta = 5$ &0.876 &0.854 &\textbf{0.821} &0.825\\
BCOM $\Delta = 21$ &0.956 &0.937 &0.874 &\textbf{0.854}\\
\hline 
INDU $\Delta = 1$ &0.451 &0.446 &\textbf{0.444} &0.458\\
INDU $\Delta = 5$ &0.617 &0.612 &\textbf{0.532} &0.541\\
INDU $\Delta = 21$ &0.858 &0.857 &0.716 &\textbf{0.699}\\
\hline 
USDEUR $\Delta = 1$ &0.530 &0.514 &\textbf{0.507} &0.521\\
USDEUR $\Delta = 5$ &0.611 &0.581 &\textbf{0.532} &0.544\\
USDEUR $\Delta = 21$ &0.755 &0.728 &\textbf{0.618} &0.638\\
\hline 
IBOV $\Delta = 1$ &0.602 &0.595 &\textbf{0.587} &0.617\\
IBOV $\Delta = 5$ &0.779 &0.754 &\textbf{0.68} &0.691\\
IBOV $\Delta = 21$ &1.010 &1.008 &0.843 &\textbf{0.836}\\
\hline 
MICROSOFT $\Delta = 1$ &0.579 &0.576 &\textbf{0.566} &0.603\\
MICROSOFT $\Delta = 5$ &0.749 &0.737 &\textbf{0.668} &0.673\\
MICROSOFT $\Delta = 21$ &0.936 &0.931 &0.807 &\textbf{0.79}\\
\hline
GOOGLE $\Delta = 1$ &0.500 &0.497 &\textbf{0.492} &0.529\\
GOOGLE $\Delta = 5$ &0.683 &0.672 &\textbf{0.581} &0.595\\
GOOGLE $\Delta = 21$ &0.864 &0.861 &0.729 &\textbf{0.722}\\
\hline 
SP400 $\Delta = 1$ &0.454 &0.451 &\textbf{0.445} &0.464\\
SP400 $\Delta = 5$ &0.616 &0.601 &\textbf{0.525} &0.538\\
SP400 $\Delta = 21$ &0.816 &0.81 &\textbf{0.668} &0.67\\
\hline
\end{longtable}

\end{center}
\noindent As we can see in Table \ref{Tab1.2}, even though RFSV sometimes underperforms AR, and HAR for $\Delta = 1$, it performs at least as good as the HAR when predicting more days ahead ($\Delta = 5, 21$) and outperforms the AR model. \\
\\
Compared to AR, HAR whose parameters change through time, and depend on the time horizon, needs to be re-calibrated and even encounters calibration issues for some periods, the RFSV is more parsimonious since it only requires the parameter $H$ to forecast the log-variance. In addition to that, the smoothness typically does not change over time or very slightly. \\
\\
We notice that prediction through the RFSV can be linked to that of \cite{DRV121}, where the issue of the prediction of the log-volatility in the multifractal random walk model of \cite{BM031} is tackled. In this model,
\begin{align*}
\mathbb{E}[\log(\sigma_{t+\Delta}^2)\mid \mathcal{F}_t] = \frac{1}{\pi}\sqrt{\Delta}\int_{-\infty}^{t}{\frac{\log(\sigma_s^2)}{(t-s+\Delta)\sqrt{t-s}}}dt,
\end{align*}
which is the limit of our predictor when $H$ tends to zero.\\
\\
The prediction formula for the RFSV model can also be rewritten as
\begin{align*}
\mathbb{E}[\log(\sigma_{t+\Delta}^2)\mid \mathcal{F}_t] = \frac{\cos(H\pi)}{\pi}\int_{0}^{+\infty}{\frac{\log(\sigma_{t-\Delta u}^2)}{(u+1)u^{H+1/2}}}du,
\end{align*}
for a given small $\varepsilon > 0$, let $r$ be the smallest real number such that
\begin{align*}
\int_{r}^{+\infty}\frac{1}{(u+1)u^{H+1/2}} \leq \varepsilon.
\end{align*}
Then we have, with an error of order $\varepsilon$
\begin{align*}
\mathbb{E}[\log(\sigma_{t+\Delta}^2)\mid \mathcal{F}_t] \approx \frac{\cos(H\pi)}{\pi}\int_{0}^{r}{\frac{\log(\sigma_{t-\Delta u}^2)}{(u+1)u^{H+1/2}}}du.
\end{align*}
This prediction formula says that future volatility depends on the whole path of the volatility process. However, since the weights decrease with time, one does not need to go to $-\infty$. It suffices to consider a time to go down to in order to forecast the future. This is roughly defined by setting the error margin $\epsilon$. For example, in order to forecast $\Delta$ in the future, it is common practice for practitioners to take $\Delta$ in the past. This corresponds to $r = 1$, and $\epsilon = 0.35$ which is not so unreasonable. 

\subsection{Predicting the variance}
\noindent Based on the same approximation of the fOU volatility process in the RFSV model, we rewrite $\sigma_t^2 = \exp(2\nu W_t^H + C)$ for some constants $\nu$ and $C$. The prediction of the variance knowing the information at time $t$ is:
\begin{align*}
\widehat{\sigma_{t+\Delta}^2} &= \mathbb{E}\left[\sigma_{t+\Delta}^2 \mid \mathcal{F}_t \right]\\
 & = \mathbb{E}\left[\exp(2\nu W_{t+\Delta}^H + C) \mid \mathcal{F}_t \right] .
\end{align*}
Since  $W_{t+\Delta}^H$ is conditionally Gaussian (as shown by  \cite{NP001} ) with conditional variance $Var[W_{t+\Delta}^H \mid \mathcal{F}_t] =  c\Delta^{2H}$ (where $c = \frac{\Gamma(3/2 - H)}{\Gamma(H+1/2)\Gamma(2-2H)}$) and using the fact that $\widehat{\log(\sigma_t^2)} \approx \mathbb{E}\left[2\nu W_t^H + C \mid \mathcal{F}_t \right]  = \mathbb{E}\left[\log(\sigma_{t+\Delta}^2) \mid \mathcal{F}_t \right]$, we have:
\begin{align*}
\widehat{\sigma_{t+\Delta}^2} = \exp \left(\widehat{\log(\sigma_{t+\Delta}^2)} + 2c\nu^2 \Delta^{2H} \right)
\end{align*}
Note that this expression uses the estimation of $\widehat{\log \left(\sigma_{t+\Delta}^2 \right)} = \mathbb{E}\left[\log(\sigma_{t+\Delta}^2) \mid \mathcal{F}_t \right]$ which we have seen in Section \ref{SectionForecastLogVol} and $\nu^2$ which is the exponential of the intercept in the linear regression of $\log(m(2,\Delta))$ on $\log(\Delta)$.\\
Once again, we compare the performance of the RFSV predictor to the AR, HAR, and the GARCH predictors expressed as the following:
\begin{itemize}
\item AR(p):
\begin{align*}
\widehat{\sigma_{t+\Delta}^2} = K_0^\Delta + \sum_{i = 0}^{p}C_i^\Delta \sigma_{t-i}^2
\end{align*}
\item HAR:
\begin{align*}
\widehat{\sigma_{t+\Delta}^2} = K_0^\Delta + C_0^\Delta \sigma_t^2 + C_5^\Delta\frac{1}{5}\sum_{i = 0}^{p}C_i^\Delta \sigma_{t-i}^2 + C_{20}^\Delta\frac{1}{20}\sum_{i=0}^{20}\sigma_{t-i}^2
\end{align*}
\item GARCH(1,1):
\begin{align*}
\widehat{\sigma_{t+\Delta}^2} = \alpha_0\left(1+ \sum_{i = 1}^{\Delta-1}{(\alpha_1 + \beta_1)^{i}}\right) + (\alpha_1 + \beta_1)^{\Delta}\sigma_t^2
\end{align*}
\end{itemize}
Results on the variance prediction are given in Table \ref{Tab2}. We can see that the RFSV model outperforms other predictors on all the considered time horizons. GARCH model performs poorly on the other hand.
\begin{table}\label{Tab2}
\centering
\small
\begin{longtable}{| p{.25\textwidth} | p{.10\textwidth} || p{.10\textwidth} | | p{.10\textwidth} | | p{.15\textwidth} | | p{.10\textwidth} |} 
Ticker  & AR(5) & AR(10) & HAR(3) & GARCH(1,1) & RFSV \\
\hline
SP100 $\Delta = 1$ &0.901 &1.01 &0.769 &0.873 &\textbf{0.655}\\
SP100 $\Delta = 5$ &1 &0.96 &1.06 &1.14 &\textbf{0.76}\\
SP100 $\Delta = 21$ &1.42 &1.33 &0.989 &1.72 &\textbf{0.898}\\
\hline
IBEX35 $\Delta = 1$ &0.62 &0.632 &\textbf{0.587} &0.675 &0.694\\
IBEX35 $\Delta = 5$ &1 &1.01 &0.846 &1.27 &\textbf{0.808}\\
IBEX35 $\Delta = 21$ &1.43 &1.45 &1.03 &2.03 &\textbf{0.975}\\
\hline
HSI $\Delta = 1$ &0.993 &1.16 &0.847 &1.37 &\textbf{0.794}\\
HSI $\Delta = 5$ &0.875 &1.16 &0.919 &1.59 &\textbf{0.851}\\
HSI $\Delta = 21$ &1.1 &1.56 &0.977 &1.81 &\textbf{0.932}\\
\hline
MEXBOL $\Delta = 1$ &0.58 &0.591 &\textbf{0.566} &0.686 &0.659\\
MEXBOL $\Delta = 5$ &0.938 &0.866 &0.808 &1.26 &\textbf{0.767}\\
MEXBOL $\Delta = 21$ &1.32 &1.31 &0.965 &1.77 &\textbf{0.928}\\
\hline
FTSE100 $\Delta = 1$ &0.65 &0.67 &\textbf{0.618} &0.776 &0.646\\
FTSE100 $\Delta = 5$ &0.833 &0.908 &0.808 &1.25 &\textbf{0.721}\\
FTSE100 $\Delta = 21$ &1.15 &1.21 &0.926 &1.69 &\textbf{0.872}\\
\hline
ASX200 $\Delta = 1$ &0.789 &0.834 &0.688 &1.08 &\textbf{0.656}\\
ASX200 $\Delta = 5$ &0.826 &0.845 &\textbf{0.721} &1.25 &0.725\\
ASX200 $\Delta = 21$ &1 &1.05 &0.851 &1.8 &\textbf{0.837}\\
\hline
TOTAL $\Delta = 1$ &0.519 &0.554 &\textbf{0.497} &0.587 &0.568\\
TOTAL $\Delta = 5$ &0.804 &0.855 &0.77 &1.01 &\textbf{0.695}\\
TOTAL $\Delta = 21$ &1.29 &1.39 &0.997 &1.59 &\textbf{0.885}\\
\hline
XIN9I $\Delta = 1$ &0.847 &0.86 &0.816 &1.24 &\textbf{0.787}\\
XIN9I $\Delta = 5$ &0.943 &0.945 &\textbf{0.86} &1.54 &0.862\\
XIN9I $\Delta = 21$ &1.1 &1.11 &0.932 &1.87 &\textbf{0.926}\\
\hline
SHSZ300 $\Delta = 1$ &0.841 &0.853 &0.806 &1.17 &\textbf{0.767}\\
SHSZ300 $\Delta = 5$ &0.964 &0.963 &0.859 &1.51 &\textbf{0.849}\\
SHSZ300 $\Delta = 21$ &1.06 &1.07 &0.918 &1.96 &\textbf{0.903}\\
\hline 
BCOM $\Delta = 1$ &0.823 &0.82 &\textbf{0.776} &1.28 &0.824\\
BCOM $\Delta = 5$ &0.884 &0.851 &0.818 &1.52 &\textbf{0.812}\\
BCOM $\Delta = 21$ &0.967 &0.961 &0.86 &1.95 &\textbf{0.834}\\
\hline
INDU $\Delta = 1$ &1.22 &1.4 &0.883 &0.938 &\textbf{0.677}\\
INDU $\Delta = 5$ &1.04 &1.05 &1.35 &1.22 &\textbf{0.779}\\
INDU $\Delta = 21$ &1.43 &1.42 &0.986 &1.76 &\textbf{0.904}\\
\hline
USDEUR $\Delta = 1$ &0.711 &0.73 &\textbf{0.663} &0.945 &0.673\\
USDEUR $\Delta = 5$ &0.878 &0.855 &0.762 &1.25 &\textbf{0.716}\\
USDEUR $\Delta = 21$ &1.05 &1.03 &\textbf{0.816} &1.46 &0.823\\
\hline
IBOV $\Delta = 1$ &0.899 &1.04 &0.687 &0.894 &\textbf{0.686}\\
IBOV $\Delta = 5$ &1.05 &0.896 &0.908 &1.11 &\textbf{0.782}\\
IBOV $\Delta = 21$ &1.74 &1.33 &1.06 &1.75 &\textbf{0.932}\\
\hline
MICROSOFT $\Delta = 1$ &0.656 &0.67 &\textbf{0.644} &0.925 &0.659\\
MICROSOFT $\Delta = 5$ &0.926 &0.894 &0.91 &1.29 &\textbf{0.746}\\
MICROSOFT $\Delta = 21$ &1.12 &1.15 &0.95 &1.61 &\textbf{0.866}\\
\hline
GOOGLE $\Delta = 1$ &0.652 &0.661 &0.611 &0.887 &\textbf{0.586}\\
GOOGLE $\Delta = 5$ &0.715 &0.737 &0.66 &1.08 &\textbf{0.651}\\
GOOGLE $\Delta = 21$ &1.09 &1.11 &0.87 &1.43 &\textbf{0.831}\\
\hline
SP400 $\Delta = 1$ &0.704 &0.761 &0.634 &0.916 &\textbf{0.626}\\
SP400 $\Delta = 5$ &0.961 &0.882 &0.778 &0.986 &\textbf{0.725}\\
SP400 $\Delta = 21$ &1.27 &1.19 &0.882 &1.56 &\textbf{0.866}\\
\hline
\end{longtable}

\end{table}
\section{Conclusion}\label{Sect1.6}
\noindent We aimed through this analysis to investigate the scaling behavior of the volatility range-based proxy. Gatheral and co-authors have already studied high-frequency-based realized-volatility in this sense in \cite{GJR141}. They found that it exhibits a rough scaling behavior and that its logarithm behaves like a fractional Brownian motion with Hurst exponent of order $0.14$. We applied their analysis using the range-based volatilities as proxies for spot volatility. The latter uses only information about range prices (open, close, high, low) for their estimation. We also find that the volatility process is monofractal with a small Hurst exponent (lower than 0.1) and can indeed be as low as $0.014$.\\ 
\\
Further tests justify that log-volatility increments are approximately Gaussian. This allows us to model them using the RFSV model. To ensure that the RFSV is consistent with the results on data, we simulate intraday and range prices using this model, recover the realized volatility, and compute again the range-based estimators. We perform again the statistical checking only to find that simulations give similar results to the data. This reinforces that the hypothesis of rough volatility can not be rejected at this point, while other models like FSV seem to be misleading and, to a certain extent, wrong. \\
\\
Finally, to measure its prediction power, we compare RFSV with other models such as AR or HAR for the log-volatility prediction and with AR, HAR, and GARCH for the variance prediction. The RFSV shows good performance compared to other models and its prediction power is at least comparable to that of the HAR. GARCH, on the other hand, shows very weak performance for these estimators.

\bibliographystyle{abbrv}
\bibliography{RoughVolatilityRB}
\end{document}